\documentclass[amsmath,
amssymb,
a4paper,
aps,prl, superscriptaddress,
twocolumn, reprint, % layout option (preprint, reprint, twocolumn)
fleqn,
showpacs,
floatfix, longbibliography]{revtex4-2}

\usepackage{amsmath}
\usepackage[table]{xcolor}
\usepackage[T1]{fontenc}
\usepackage{lmodern}
\usepackage{bbm}
\usepackage{microtype}
\usepackage{mathtools}%, cuted}
\usepackage{mathcomp}
\usepackage{color}
\usepackage{tabularx}
\usepackage{stmaryrd}
\usepackage{booktabs}
\usepackage{multirow}

\usepackage{xcolor}
\usepackage[
format=hang,
indention=-1.0cm,
justification=RaggedRight,
font=scriptsize
]{caption}

\usepackage{graphicx} % Include figure files
\usepackage{dcolumn} % Align table columns on decimal point
\usepackage{bm} % bold math
\usepackage[version=4]{mhchem}
\usepackage{stackrel}
\usepackage{verbatim}   
\usepackage{hyperref}

\usepackage[geometry]{ifsym}
\usepackage[utf8]{inputenc}
\usepackage[english]{babel}
\usepackage{calc}
\usepackage{graphicx}
\usepackage{textgreek}

\graphicspath{{Figures/}} 

\newlength\myheight
\newlength\mydepth
\settototalheight\myheight{Xygp}
\newcommand{\be}{\begin{equation}}
\newcommand{\ee}{\end{equation}}
\newcommand{\bea}{\begin{eqnarray}}
\newcommand{\eea}{\end{eqnarray}}

\newcommand{\ffor}{\mathrel \triangleright \joinrel \mathrel { \triangleright}}

\newcommand{\copro}{\text{\footnotesize{$\Upsilon$}}}

\newcommand{\mathleft}{\@fleqntrue\@mathmargin0pt}
\newcommand{\mathcenter}{\@fleqnfalse}

\usepackage{academicons}
\definecolor{orcidlogocol}{HTML}{A6CE39}

\renewcommand{\selectlanguage}[1]{}
\usepackage{orcidlink}

\begin{document}

\title{Emergent conserved quantities via irreversibility} %emergent conservation laws through irreversibility} %On conserved quantities in reaction networks due to irreversible reactions
\author{Alex Blokhuis\orcidlink{0000-0002-4594-596X}}
\email{alexander.willem@imdea.org}
\affiliation{Stratingh Institute for Chemistry, University of Groningen, Nijenborgh 4, 9747 AG Groningen, the Netherlands}
\affiliation{Groningen Institute for Evolutionary Life Sciences, University of Groningen, Nijenborgh 4, 9747 AG Groningen, the Netherlands}
\affiliation{University of Strasbourg $\&$ CNRS, UMR7140, 67083 Strasbourg, France} 
\affiliation{Instituto IMDEA Nanociencia, Calle Faraday 9, 28049 Madrid, Spain} 
\author{Martijn van Kuppeveld\orcidlink{0000-0001-9045-9322}}\affiliation{Radboud Center for Natural Philosophy, Radboud University Nijmegen, Heyendaalseweg
135, 6525 AJ Nijmegen, The Netherlands}
\affiliation{Stratingh Institute for Chemistry, University of Groningen, Nijenborgh 4, 9747 AG Groningen, the Netherlands}

\author{Daan van de Weem}\affiliation{Scuola Internazionale Superiore di Studi Avanzati (SISSA), Via Bonomea 265, 34136 Trieste, Italy}

\author{Robert Pollice\orcidlink{0000-0001-8836-6266}}
\email{r.pollice@rug.nl}
\affiliation{Stratingh Institute for Chemistry, University of Groningen, Nijenborgh 4, 9747 AG Groningen, the Netherlands}

\date{\today} 
%===================
\begin{abstract} 
Conserved quantities increasingly underpin the inference of physical models. Recently new conserved quantities have been found in this context, that currently lack an interpretation. Here, we show that irreversible reactions in CRNs and Markov Chains lead to emergent conservation laws and broken cycles. Linearly dependent currents - characterized by the "co-production index" - arise due to irreversible reactions. We derive a law relating conserved quantities, broken cycles, and co-production. This resolves a recent conundrum posed by a machine-discovered candidate for a non-integer conservation law. Our findings introduce heretofore overlooked extensions to a widely used index law for CRNs and Markov Chains that undercounts conservation laws. This furnishes new tools and immediate applications for the inference and analysis of models based on conservation laws.
\end{abstract}

\maketitle

\begin{figure}[tbhp!]
\centering
\includegraphics[width=1.0\linewidth]{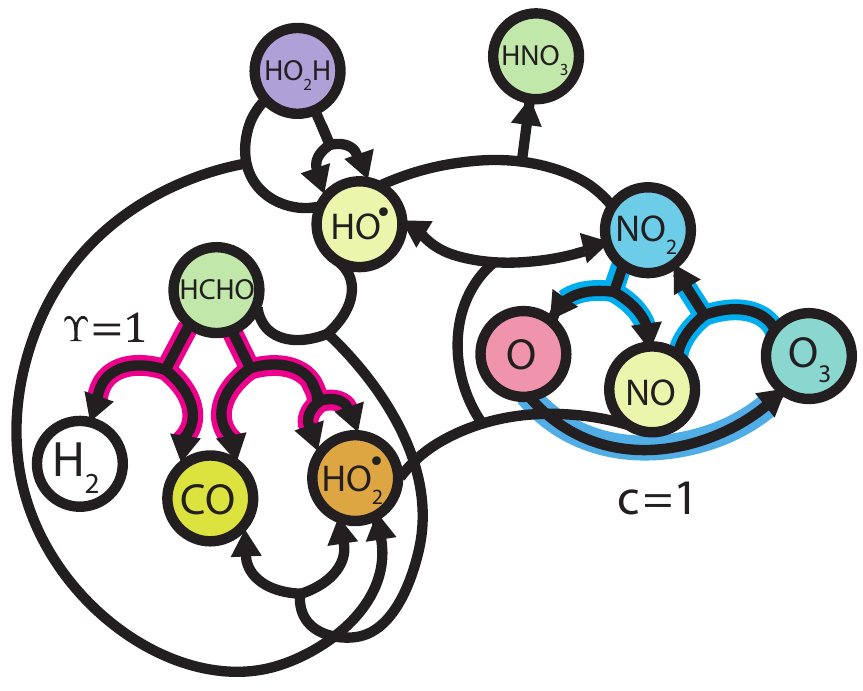}
\caption{\justifying An Atmospheric CRN model\cite{Sturm2020,Sturm2022} with a recent machine-discovered candidate conserved quantity ($\text{CQ}_3$ in Ref.\cite{Liu_PRE_2024}). The co-production index $\text{\footnotesize{$\Upsilon$}}=1$ is given next to its two collinear reactions (also highlighted in pink). These do not form a cycle. The cycle index $c=1$ is given next to its three cycle reactions (highlighted in blue). Thus, no cycles break upon merging collinear reactions: $\Delta c = 0$, $\Delta \ell = +1$, i.e., a co-production conservation law then results from the co-production law $\text{\footnotesize{$\Upsilon$}}= \Delta \ell - \Delta c$, confirming the numerically found $\text{CQ}_3$ marks the detection of a genuine conservation law.}
\label{fig:atmosphericnetwork}
\end{figure}

\textit{Introduction---}Our capacity to build \cite{kriukov_history_2022,ivanov_computing_2023,zhang_complex_2018,sharko_redox-controlled_2023,singh_devising_2020,semenov_rational_2015,meijer_hierarchical_2017,Chen2023,ikeda_installing_2014,donau_phase_2022} and understand\cite{aris_prolegomena_1965,polettini_irreversible_2014,feinberg_foundations_2019,sughiyama_chemical_2022,dal_cengio_geometry_2023,avanzini_circuit_2023,Schnakenberg1976,hill_free_1989,baez2018biochemical,baez_compositional_2017,blokhuis_universal_2020,Vassena2024,deshpande_autocatalysis_2014,andersen_defining_2021,avanzini2023methods,Zhang2023_FT} chemical systems of increasing complexity is intimately tied to our understanding of their reaction networks and our ability to elucidate them. Whereas sizeable molecular structures can be readily elucidated today, reconstructing the structure of even small reaction networks remains challenging\cite{unsleber_exploration_2020}. Reciprocally, deducing from structure alone whether a CRN can exhibit nontrivial behavior (e.g., oscillations) remains largely open (but see \cite{stoichrec2025,blokhuis_universal_2020}). Strikingly, predicting the total number of observable conserved quantities is also still open: classical laws predicting the number of conserved quantities\cite{aris_prolegomena_1965,polettini_irreversible_2014} are found to undercount conserved quantities in several instances\cite{Liu_PRE_2024,Liu2021}. This poses a clear obstacle for efforts in machine learning (ML) and big data rooted in finding conserved quantities with a physical interpretation\cite{Liu2021,Li2023_metalearning,Liu2022,Beucler2021,Ha2021,Liu2022_diffeq,Mebratie2025}, as exemplified by a recent ML work\cite{Liu_PRE_2024} that found a possible anomalous conserved quantity that could not be theoretically accounted for from dynamics in an atmospheric chemistry model \cite{Sturm2020,Sturm2022} (cf. Fig. \ref{fig:atmosphericnetwork}). In this letter, we provide an independent extension of the first structural law whereby such conservation laws can be counted and interpreted in terms of collinear irreversible reactions: co-production.

\textit{Setup---}

We consider chemical reaction networks (CRNs, with Markov chains as special linear case), with reactions given by
\be
\sum_k \mathbb{S}^{\ominus}_{k,i} \ce{X_k} \overset{i}{\rightarrow} \sum_k \mathbb{S}^{\oplus}_{k,i} \ce{X_k}
\ee 
where $\mathbb{S}^{\ominus}_{k,i}$ (resp. $ \mathbb{S}^{\oplus}_{k,i}$) are stoichiometric coefficients for reactant (resp. product) species $\ce{X}_k$ in reaction $i$. Overall reaction stoichiometry can be written as $s$-by-$r$ stoichiometric matrix $\mathbb{S}$ as $\mathbb{S} = \mathbb{S}^{\oplus} - \mathbb{S}^{\ominus}$. 

Dimensions of fundamental subspaces of this matrix - image, kernel, co-image, co-kernel - count physically meaningful quantities. Through the fundamental theorem of linear algebra, a key relation we will call the 'first structural law' (SL1) can then be formulated
\be
\text{rk}(\mathbb{S}) = s - \ell = r - c \label{equation:SL1}
\ee
denoting $\#$  'the number of', we normally have: \\
{$\text{rk}(\mathbb{S})$}: rank of matrix $\mathbb{S}$ \\
{$s: \ \#$} species ({$\ce{X_1}, ... , \ce{X_s}$})\\
{$\ell: \ \#$} "standard" conserved quantities\footnote{with the caveat that these are normally undercounted, and this paper supplements remaining ones} {$(\text{dim}(\text{coker}(\mathbb{S}))$} \\
{$r: \ \#$} reactions ({$r_1, ..., r_r$}) \\
{$c: \ \#$} cycles {$(\text{dim}(\text{ker}(\mathbb{S})))$} \\
Cycles correspond to combinations of reactions (transitions), in the mathematical sense, that leave the system unchanged. A conserved quantity corresponds to a linear combination of species abundances which is fixed. Symbolically,
\bea
\mathbb{S} \ \pmb{c}^{(i)} = \pmb{0} \ \ \ (i \in \{1, ..., c\}), \\
\mathbb{S}^T \pmb{\ell}^{(j)}  = \pmb{0} \ \ \ (j \in \{1, ..., \ell\}).  \label{equation:nullveceq}
\eea
In general, time evolution of species concentrations $[\ce{X_1}], ..., [\ce{X_s}]$ follows the kinetic equations
\bea
d_t [\pmb{\ce{X}}] &=& \mathbb{S} \pmb{J} , \label{equation:timeevo} \\
J_i &=& \kappa_i \prod_{k=1}^s \mathbb{S}^{\ominus}_{k,i} [\ce{X_k}]^{\mathbb{S}^{\ominus}_{k,i}}
\eea
under mass-action (see SM IV for a generalization), where $\kappa_i$ is a rate constant and $\pmb{J}
=(J_1,..,J_r)$ a vector of currents. 

\textit{Conserved quantities---}A conserved quantity $L^{(j)}$ is defined as 
\bea
&\ell_1^{(j)} [\ce{X}_1] + ... + \ell_s^{(j)} [\ce{X}_s] = L^{(j)}, \\
&d_t L^{(j)} = 0. \label{equation:dtLj}  
\eea
Rewriting Eq.\eqref{equation:dtLj} in vector form, we get
\be
 \! \! \! \! \! \! \! \! \!  d_t L^{(j)} = d_t \left(   [\pmb{\ce{X}}]^T \pmb{\ell}^{(j)} \right) =\  \pmb{J}^T \mathbb{S}^T  \pmb{\ell}^{(j)}   = 0. \label{equation:dyneq} 
\ee
The $\ell$ left nullvectors of $\mathbb{S}$, for which   $\left(\mathbb{S}^T \pmb{\ell}\right)^T = \pmb{\ell}^T  \mathbb{S} = \pmb{0}^T$, form a solution to Eq. \eqref{equation:dyneq}. While often assumed, it does \underline{not} follow that these are \underline{all} the solutions. If there are currents in $\pmb{J}$ that are linearly dependent for all $[\pmb{\ce{X}}]$\footnote{In mass action models currents are proportional to a single monomial in [X]. Therefore a linear relation among currents for all $[\pmb{\ce{X}}]$ implies that the monomials are the same with a different rate constant. For this reason all relations can be built up from pairwise relations.}, these can give rise to additional solutions of Eq.\eqref{equation:dyneq}, which would mean additional conserved quantities. In order to classify these, this paper introduces the notion of merged reactions, which gives rise to an effective stochiometric matrix $\mathbb{S}^*$ with independent effective currents, from which the previously hidden conservation laws arise as in \eqref{equation:nullveceq}, thus elucidating the full set of linear conserved quantities.

\textit{Merging reactions---}Consider a pair of irreversible reactions $r_i, r_j$ with \textit{pairwise proportional} reaction rates: $J_i \propto J_j$. We can write a \textit{merged} current $J_{*}$:
\be
J_{*} = \frac{\kappa_i}{\kappa_i+\kappa_j} J_i + \frac{\kappa_j}{\kappa_i+\kappa_j} J_i 
\ee
and a merged stoichiometry
\bea
\mathbb{S}^{\oplus}_{k,*} &=& \frac{\kappa_i}{\kappa_i+\kappa_j} \mathbb{S}^{\oplus}_{k,i} + \frac{\kappa_i}{\kappa_i+\kappa_j} \mathbb{S}^{\oplus}_{k,i}. \\
\mathbb{S}^{\ominus}_{k,*} &=& \mathbb{S}^{\ominus}_{k,i} = \mathbb{S}^{\ominus}_{k,j},
\eea
i.e., we describe a pair of pairwise proportional reactions as a single reaction. For example: 
\bea
 \! \! \! \! \! \! \! \! \! \ce{C} \overset{1}{\leftarrow} \ce{A}&+&\ce{B} \overset{2}{\rightarrow} \ce{D} \label{equation:examplecrn} \\
 \! \! \! \! \! \! \! \! \! &\Downarrow& \ merge \nonumber \\
 \! \! \! \! \! \! \! \! \! \ce{A} + \ce{B} &\overset{*}{\rightarrow}& \frac{\kappa_1}{\kappa_1 + \kappa_2} \ce{C} + \frac{\kappa_2}{\kappa_1 + \kappa_2} \ce{D}  
\eea
with a single current
\be
J_* = J_1 + J_2 = \left(\kappa_1 + \kappa_2\right) [\ce{A}] [\ce{B}] \nonumber
\ee
The corresponding stoichiometric matrices are
\be
 \! \! \! \! \! \! \! \! \!   \mathbb{S} = \left(\begin{smallmatrix} -1 & -1 \\ 
-1 & -1 \\
1 & 0 \\
0 & 1 \end{smallmatrix}\right), \ \ \mathbb{S}^{*} = \left(\begin{smallmatrix} -1  \\ 
-1  \\
\frac{\kappa_1}{\kappa_1 + \kappa_2}  \\
\frac{\kappa_2}{\kappa_1 + \kappa_2}  \end{smallmatrix}\right) = \left(\begin{smallmatrix} -1  \\ 
-1  \\
p  \\
1-p  \end{smallmatrix}\right)  \  {\textcolor{teal}{\begin{smallmatrix}  \ce{A} \\ \ce{B} \\ \ce{C} \\ \ce{D}   \end{smallmatrix}}} \nonumber %\label{equation:matrixmerge}
\ee
where successive rows act on $\ce{A}, \ce{B},\ce{C}$, and $\ce{D}$, respectively. $\mathbb{S}$ and $\mathbb{S}^*$ describe ODE evolution equivalently as
\be
d_t ([\ce{A}], [\ce{B}], [\ce{C}], [\ce{D}])^T = \mathbb{S} \begin{pmatrix} J_1 \\ J_2 \end{pmatrix} = \mathbb{S}^* J_*. 
\ee
Merged stoichiometric matrix $\mathbb{S}^*$ has fully independent currents, and $\ell^*  = 3$ conservation laws. $\mathbb{S}$ only reveals $\ell=2$ of them and misses
\bea
L^* = (1-p) [\ce{C}] + p [\ce{D}], \\
d_t L^* = (1-p) J_1 + p J_2 = 0
\eea
which admits non-integer coefficients, a distinguishing feature of such conserved quantities. 

\textit{Co-production law---}To generalize to a law, we denote co-production $\copro$ the number of times this procedure can be repeated in a CRN until no pairwise proportional reactions remain (see also SM I). Applying SL1 to resulting matrix $\mathbb{S}^*$, we have\footnote{Each pairwise merger reduces the $\#$ reactions by one. Mergers do not affect the $\#$ rows, i.e., leave species unaffected.}, using $\Delta s = 0$ 
\bea
\! \! \! \! \text{rk}(\mathbb{S}^*) - \text{rk}(\mathbb{S}) &=& \Delta s - \Delta \ell = \Delta r - \Delta c, \nonumber \\
\copro &\equiv& -\Delta r = \Delta \ell - \Delta c . \label{equation:coprolaw}
\eea
i.e., each merger either makes a conservation law emerge or breaks a cycle. \footnote{This formulation includes $\rho_{\bowtie}$ lost 2-cycles due to merging reversible reactions, which one ordinarily (e.g., in thermodynamics) does not want to include among cycles. Formal procedures for subtracting them can be found in ref. \cite{blokhuis2025datadimensionchemistry} and will be developed elsewhere.} 

\textit{Applications of co-production---} 
Co-production readily occurs where reactions are plentiful and irreversible\cite{blokhuis2024ejoc}, as found in atmospheric, astro-, photo- and radiochemistry. As reversible reactions can effectively behave as irreversible on short timescales they can give rise to transient conserved quantities (which has been used in \cite{blokhuis2024ejoc} to elucidate network structure from chemical data). The co-production law applies to Markov models as well. A master equation with $s$ states and $r$ transitions can naturally be represented using a stoichiometric matrix and currents depending on occupation probabilities $\pmb{P} = (P_1,..,P_s)^T$ (SM VIa). 
\be
d_t \pmb{P} = \mathbb{S} \pmb{J} (\pmb{P}) = \mathbb{S}^* \pmb{J}^* (\pmb{P}) 
\ee
For multispecies Markov Processes with first-order transitions one can derive a population number representation (second quantization, see also e.g.\cite{Peliti1985,Doi1976a,Doi1976b}), which affords a stoichiometric matrix representation in terms of population means $\langle \pmb{n} \rangle = (\langle n_1 \rangle , .., \langle n_s \rangle )^T$ (SM VIb)
\be
d_t \langle \pmb{n} \rangle = \mathbb{S} \pmb{J} (\langle \pmb{n} \rangle) = \mathbb{S}^* \pmb{J}^* (\langle \pmb{n} \rangle)
\ee
Such a description describes a wide variety of models in satistical physics, e.g. for adsorbtion, fragmentation, asset-exchange, diffusion and can readily reveal hidden conserved quantities. The rest of this paper will be dedicated to demonstrating two applications of the co-production law \eqref{equation:coprolaw}, one in atmospheric chemistry and one in Markov models.

%by mapping state populations (e.g.$[k]$) to chemical species. 
%\bea
 %\! \! \! \! \! \! \! \! \!  \langle [k] \rangle_{m} &=& \langle [k] \rangle_{m-1} +  \mathbb{S} \pmb{J}_{m-1} \ \ \ (\text{discrete})     \\
 % \! \! \! \! \! \! \! \! \! \pmb{J}_{m-1} &=& \text{diag}(\pmb{\kappa}) \langle [k] \rangle_{m-1} \nonumber \\
 %\! \! \! \! \! \! \! \! \!  d_t \langle [k] \rangle &=& \mathbb{S} \pmb{J}  \ \ \ \ \ \ \ \ \ \ \ \ \ \ \ \ \ \ \  (\text{continuous}) \\
 %\! \! \! \! \! \! \! \! \! \pmb{J} &=& \text{diag}(\pmb{\kappa}) \langle [k] \rangle  \nonumber 
%\eea

%Co-production readily occurs where reactions are plentiful and irreversible\cite{blokhuis2024ejoc}, as found in  astro-, photo- and radiochemistry, as well as in atmospheric chemistry, for instance in the next example. Reversible reactions can transiently behave as irreversible and \textit{ipso facto} manifest as (transient) conserved quantities in experimental data informative of network structure\cite{blokhuis2024ejoc}. Finally, Markov models can be mapped to linear CRNs by mapping states to species. 

\textit{Revisiting a machine-discovered law---}From this law, we can now immediately corroborate that a recent machine-discovered candidate conservation law is indeed a valid conservation law, as seen in Fig. \ref{fig:atmosphericnetwork} and analyzed in detail in sec. II of the Supp. Matt. In short: a co-production index $\copro = 1$ follows from visual inspection of collinear reactions in Fig. \ref{fig:atmosphericnetwork} (highlighted next to $\copro=1$) or algorithmically from the stoichiometric matrix $\mathbb{S}$.

$\mathbb{S}$ is then given by ($r_4$, $r_5$ are collinear)
\hspace*{-0.75cm}\vbox{\be
\mathbb{S} = \left(\begin{smallmatrix} 0 & 1 & -1 & 0 & 0 & 0 & 0 & 0 & 0 & 0 \\
1 & 0 & -1 & 0 & 0 & 0 & -1 & 0 & 0 & 0  \\
-1 & 0 & 1 & 0 & 0 & 0 & 1 & -1 & 0 & 0 \\
0 & 0 & 0 & \colorbox{pink}{-1} & \colorbox{pink}{-1} & -1 & 0 & 0 & 0 & 0 \\
0 & 0 & 0 & \colorbox{pink}{2} & 0 & 1 & -1 & 0  & 0 & 1 \\
0 & 0 & 0 & 0 & 0 & 0 & 0 & 0 & -1 & -1 \\
0 & 0 & 0 & 0 & 0 & -1 & 1 & -1 & 2 & -1 \\
1 & -1 & 0 & 0 & 0 & 0 & 0 & 0 & 0 & 0 \\
0 & 0 & 0 & 0 & 0 & 0 & 0 & 1 & 0 & 0 \\
0 & 0 & 0 & \colorbox{pink}{1} & \colorbox{pink}{1} & 1 & 0 & 0 & 0 & 0 \\
0 & 0 & 0 & 0 & \colorbox{pink}{1} & 0 & 0 & 0 & 0 & 0
\end{smallmatrix}\right)   \ \ \ {\textcolor{teal}{\begin{smallmatrix}  \ce{O3} \\  \ce{NO} \\ \ce{NO2} \\ \ce{HCHO} \\ \ce{HO2} \\ \ce{HO2H} \\ \ce{OH} \\ \ce{O} \\ \ce{HNO3} \\ \ce{CO} \\ \ce{H2}   \end{smallmatrix}}} \ \ .
\ee}

By merging $r_4$ and $r_5$, a stoichiometric matrix $\mathbb{S}^{*}$ of independent reactions is obtained:
\hspace*{-0.75cm}\vbox{\be
\mathbb{S}^{*} = \left(\begin{smallmatrix} 
0 & 1 & -1 & 0 & 0 & 0 & 0 & 0 & 0 \\
1 & 0 & -1 & 0 & 0 & -1 & 0 & 0 & 0  \\
-1 & 0 & 1 & 0 & 0 & 1 & -1 & 0 & 0 \\
0 & 0 & 0 & \colorbox{pink}{-1} & -1 & 0 & 0 & 0 & 0 \\
0 & 0 & 0 & \colorbox{pink}{$2 p$} & 1 & -1 & 0  & 0 & 1 \\
0 & 0 & 0 & 0  & 0 & 0 & 0 & -1 & -1 \\
0 & 0 & 0 & 0  & -1 & 1 & -1 & 2 & -1 \\
1 & -1 & 0 & 0  & 0 & 0 & 0 & 0 & 0 \\
0 & 0 & 0 & 0  & 0 & 0 & 1 & 0 & 0 \\
0 & 0 & 0 & \colorbox{pink}{1}  & 1 & 0 & 0 & 0 & 0 \\
0 & 0 & 0 & \colorbox{pink}{1-$p$} & 0 & 0 & 0 & 0 & 0 \end{smallmatrix}\right).
\ee}

From this, a new co-production law emerges as left nullvector:
\hspace*{-0.75cm}\vbox{\be
\pmb{\ell}_{*} = \left(6,-5,1,3,9,6,3,6,4,-3,\frac{6 - 18 p}{1-p} \right)^T . 
\ee}
We show in SM II that this is the quantity found by the SID algorithm in Ref. \cite{Liu_PRE_2024}. Co-production thus confirms the validity of this heretofore unexplained numerical result found through ML, and provides the missing theoretical foundation to interpret it.

\textit{Emergent conservation laws in Markov processes---}
As currents in Markov models (in CRN form) are linear functions, co-production is a generic feature in Markov models, which turn out to exhibit many conserved quantities beyond conservation of total probability. We illustrate this point by applying it to n-mer adsorption and provide two further examples of co-production in diffusion models in SM III. 

\begin{figure}[tbhp!]
\centering
\includegraphics[width=1.0\linewidth]{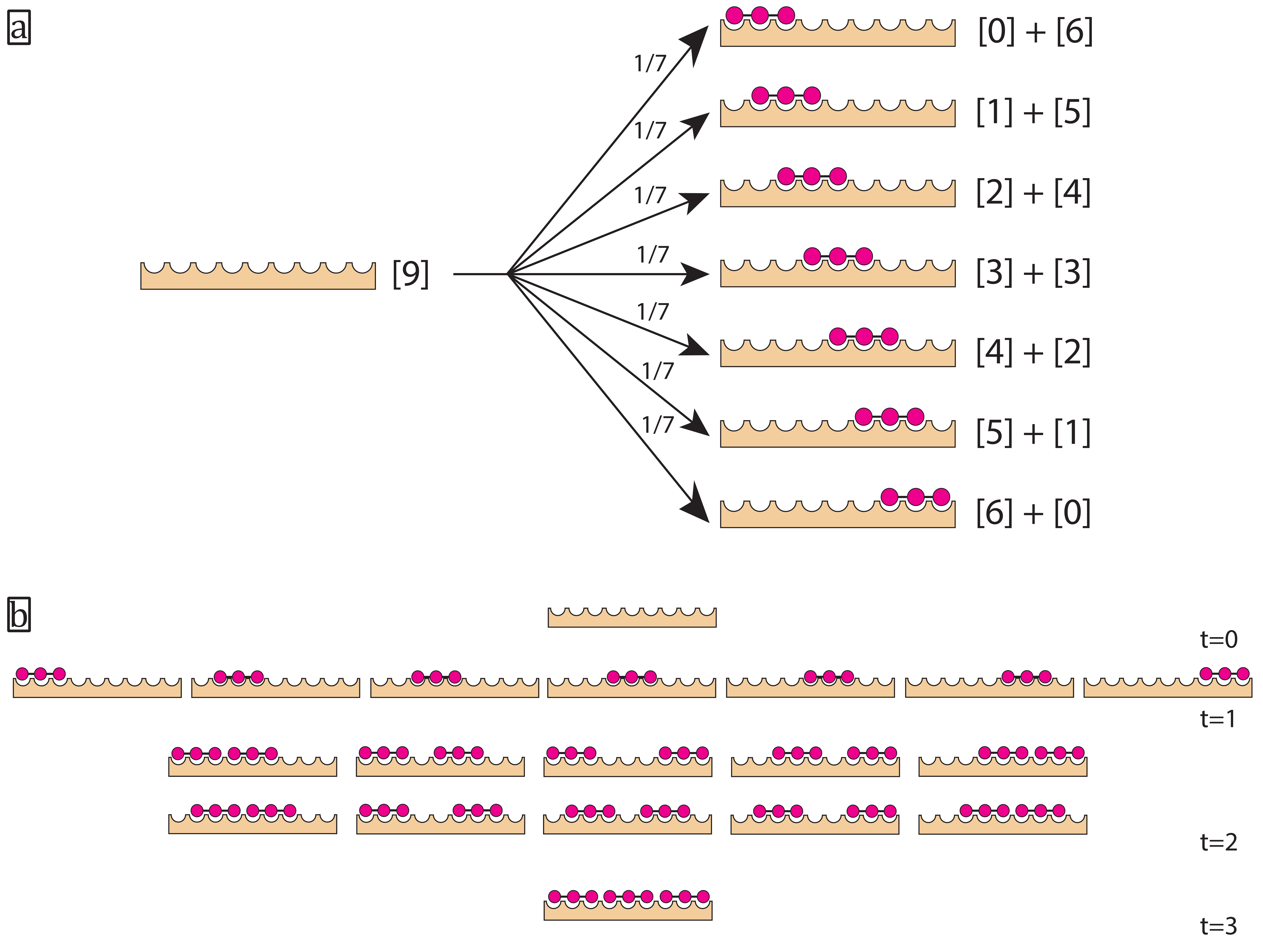}
\caption{a. Adsorbtions for trimers ($n=3$) in a random sequential adsorption model, lattice size $\lambda=9$. b) Configurations after 0, 1, 2, and 3 steps. Most configurations are jammed after 2 steps.}
\label{fig:orbtion}
\end{figure}

We consider the irreversible random sequential adsorption\cite{Evans1993,Krapivsky2010} (RSA\footnote{See Ref.\cite{Evans1993} for a review on discrete RSA models and Ref. \cite{Talbot2000} for a review on continuous RSA models.}) of n-mers on an initially empty discrete lattice of $\lambda$ sites . We denote with $[k]$ the population of gaps of k successive empty sites, which an n-mer adsorption splits in two gap populations (Fig.\ref{fig:orbtion}a)
\hspace*{-0.5cm}\vbox{\be
\ \! \! \! [m+n] \rightarrow [m-k] + [k] \ \ \ (0 \leq k \leq m) .
\ee}
When merging collinear reactions, we have
\be
\ \! \! \! [m+n] \rightarrow \frac{1}{m+1} \sum_{k=0}^{m} \left( [m-k] + [k] \right),  \nonumber 
\ee
which simplifies to
\be
\ \! \! \! [m+n] \rightarrow \frac{2}{m+1} \sum_{k=0}^{m} [k] ,
\ee
which, in a stoichiometric matrix, yield columns of the form $(-1, [0]_n, [\frac{2}{m+1}]_{m+1}$). For $\lambda=9$, $n=3$, we have the stoichiometric matrix\footnote{Columns for $[7], \;[8]$ can be omitted for $\lambda=9, \;n=3$, since these populations cannot be changed. They are included here to better illustrate the regular structure of $\mathbb{S}_*$} 
\hspace*{-0.75cm}\vbox{\be  \! \!  \! \!
\mathbb{S}^{*} = \begin{pmatrix}
\frac{2}{7} & \frac{2}{6} & \frac{2}{5} & \frac{2}{4} & \frac{2}{3} & 1 & 2 \\

\frac{2}{7} & \frac{2}{6} & \frac{2}{5} & \frac{2}{4} & \frac{2}{3} & 1 & 0  \\		  

\frac{2}{7} & \frac{2}{6} & \frac{2}{5} & \frac{2}{4} & \frac{2}{3} & 0 & 0  \\

\frac{2}{7} & \frac{2}{6} & \frac{2}{5} & \frac{2}{4} & 0 & 0 & -1 \\
          
\frac{2}{7} & \frac{2}{6} & \frac{2}{5} & 0 & 0 & -1 & 0 \\ 

\frac{2}{7} & \frac{2}{6} & 0 & 0 & -1 & 0 & 0 \\

\frac{2}{7} & 0 & 0 & -1 & 0 & 0 & 0 \\

0  & 0 & -1 & 0 & 0 & 0 & 0  \\  

0  & -1 & 0 & 0 & 0 & 0 & 0  \\  

-1  & 0 & 0 & 0 & 0 & 0 & 0  
      \end{pmatrix}   \ \ \ \textcolor{teal}{\begin{matrix} [0] \\ [1] \\ [2] \\ [3] \\ [4] \\ [5] \\ [6] \\ [7] \\ [8] \\ [9] \end{matrix}} \ .  
\ee}
with currents $\pmb{J} = (J_1,..,J_7)^T$ described by
\be
J_k = \kappa_k \langle k+2 \rangle , \ \ \ \kappa_k = k \kappa 
\ee
Gaps smaller than $n$ (here: $[2], \; [1], \; [0]$) cannot be reduced further, gaps of size $n$ or greater always have an outgoing reaction. 
We obtain $n$ co-production conservation laws $\pmb{\ell}^{(0)}_*,..,\pmb{\ell}^{(n-1)}_*$ governing state probabilities (see Eq. \eqref{equation:gensol_nmer} for a general solution), satisfying $ \mathbb{S}_{*}^T \pmb{\ell}_* = \pmb{0}$, yielding a conservation law on averages:
\be
\! \! \! \! L^{(k)}_{n,\lambda} = \pmb{\ell}^{(k)} \cdot \begin{pmatrix} \langle [0]_t \rangle, \langle [1]_t \rangle,
\dots  \langle [\lambda]_t \rangle  \end{pmatrix}^T . \label{equation:cons_avg_nmer}
\ee

A convenient basis for these conservation laws is one where each vector has a nonzero argument for only a single of the irreducible gaps:
\bea
\! \! \! \! \pmb{\ell}^{(0)} &=& \left( 1, 0, 0, 2, 1, \frac{2}{3}, \frac{3}{2}, \frac{8}{5}, \frac{14}{9}, \frac{37}{21} \right)^T , \nonumber \\
\! \! \! \! \pmb{\ell}^{(1)} &=& \left( 0, 1, 0, 0, 1, \frac{6}{9}, \frac{1}{2}, \frac{4}{5}, \frac{8}{9}, \frac{19}{21} \right)^T ,
\nonumber \\ 
\! \! \! \! \pmb{\ell}^{(2)} &=& \left(0,0,1,0,0,\frac{6}{9},\frac{1}{2},\frac{2}{5},\frac{5}{9},\frac{13}{21} \right)^T. \label{equation:coeffcons_law_nmer}
\eea

$L^{(0)}, \; L^{(1)}, \; L^{(2)}$ now follow from initial condition $[L]_0 = 1, \; [k]_0 = 0 \; (k\neq \lambda) $ for empty lattice,
\bea
L^{(j)}_{n,\lambda} &=& \ell^{(j)}_\lambda [\lambda]_0 = \ell^{(j)}_\lambda, \ \ \  \nonumber \\
L^{(1)}_{3,9} &=& \frac{37}{21}, \ \ \ L^{(2)}_{3,9} = \frac{19}{21}, \ \ \ L^{(3)}_{3,9} = \frac{13}{21}. \nonumber
\eea

Evaluation at $t\rightarrow\infty$ yields exact gap populations $\langle [k]_\infty\rangle$ and jamming coverages $\rho_{n,\lambda}$, the fraction of sites covered when no further adsorbtion can occur:
\bea
\langle [k] \rangle_{\infty} &=& L^{(k)},  \\
\rho_{n,\lambda} &=& 1 - \sum_{k=0}^{n-1} \frac{k L^{(k)}}{\lambda} .
\eea

In SM III, we derive the following algebraic expression for coefficients $\ell_{j}^{(q)}$ in $L^{(q)}_{n,\lambda}$:
\hspace*{-1.0cm}\vbox{\bea
&\begin{pmatrix} \ell^{(q)}_{mn-n} \\ \ell^{(q)}_{mn-n+1} \\ \vdots \\ \ell^{(q)}_{mn-1}  \end{pmatrix} = \prod_{p=1}^{m-1} \mathbb{F}_{p,n} \begin{pmatrix}\ell^{(q)}_0 \\ \ell^{(q)}_{1} \\ \vdots  \\ \ell^{(q)}_{n-1} \end{pmatrix} \label{equation:gensol_nmerb} \\ \vspace*{0.5cm}
&\mathbb{F}_{p,n} = \left(\begin{smallmatrix}
\frac{2}{np-n+1} & 0 & 0  & \dots & 0 &    \frac{np-n}{np-n+1} \\
 \frac{2}{np-n+2} & \frac{2}{np-n+2} & 0 &  \dots & 0 & \frac{np-n}{np-n+2} \\
\vdots & \vdots & \vdots & \ddots & \vdots & \vdots \\
 \frac{2}{np-2} & \frac{2}{np-2} & \frac{2}{np-2}  & \dots & 0 & \frac{np-n}{np-2} \\
 \frac{2}{np-1}  & \frac{2}{np-1} & \frac{2}{np-1} & \dots & \frac{2}{np-1}  & \frac{np-n}{np-1} \\
 \frac{2}{np} & \frac{2}{np} &  \frac{2}{np}  & \dots  & \frac{2}{np} & \frac{np-n+2}{np} 
\end{smallmatrix} \right) . \nonumber
  \eea}
  
In SM III, we illustrate how to approach more complicated Markov Processes, e.g., RSA with mixtures of oligomers and derive analytical solutions and conservation laws for them through co-production.

\textit{Discussion---}
Co-production has observable experimental consequences, in particular in the dimensionality of chemical data and the occurrence of features like isosbestic points\cite{blokhuis2024ejoc}. Reciprocally, co-production is pertinent for the correct interpretation of dimensional observations in data. Specifically, the structural law of CRN theory - which counts conservation laws - does not include co-production. In Ref. \cite{Liu_PRE_2024}, this led to the report of a conservation law in an atmospheric chemical network that could not be accounted for. Counting the correct number of (linear) conserved quantities requires co-production, which has important implications for CRN extraction and ML. It turns out that the number of overlooked conservation laws can be considerable, especially for chemistry with many irreversible reactions (e.g., atmospheric, photo- or radiochemistry) or Markov processes. 
The inclusion of a previously overlooked conservation law also furnishes a theoretical tool for analysis, which we illustrated for the random sequential adsorption model. From co-production, we readily obtained conservation laws applicable throughout the process and analytical expressions of saturation coverages and gap populations for arbitrary sizes of lattice and k-mers (or mixtures thereof). 

Stochastic thermodynamics has recently started to unearth current-current relations, including affine\cite{Harunari2024,Bebon2026,Khodabandehlou_2025} relations and their multilinear generalizations\cite{cengio2025,Polettini2026} among stationary nonequilibrium currents in Markov Networks. Co-production provides a new body of results beyond Markov processes and stationary states to the developing corpus of current-current relations. We surmise that these relations are connected and that there are deeper current-current relations yet to be elucidated, along with dimensional consequences. 

\textit{Conclusion---}We demonstrated emergent conservation laws and broken cycles that arise as structural effects of irreversible reactions (or transitions) in CRNs and Markov processes. While reactions are classically treated as independent, this is not entirely the case: pairs of irreversible reactions can have pairwise proportional rates and then effectively behave as a single reaction (co-production). This either results in a conservation law or a broken cycle, as captured by the co-production law, which extends a central structural law in CRN theory. 

In short, the co-production law extends the toolbox of statistical physics and the corpus of current-current relations\cite{Harunari2024,cengio2025,Polettini2026} in stochastic thermodynamics and lays a foundation towards the first-principles, physics-based inference of models from data\cite{blokhuis2024ejoc}.

\textit{Acknowledgements---} A.B. acknowledges support from the EU (Marie-Sklodowska-Curie grant 847675).
A.B. acknowledges fruitful discussions with Sijbren Otto.
We thank Troy Figiel for drawing our attention to the unexplained conserved quantity in Ref. \cite{Liu_PRE_2024}.

%Our law opens the door to 

%Overall, we have introduced a fundamental law 

%\bibliographystyle{apsrev4-2.bst}
\bibliography{BibFile}

\section{Appendix}

\section{General procedure for merging reactions}
\label{subsection:genprocmerg}

\begin{figure}[tbhp!]
\centering
\includegraphics[width=1.0\linewidth]{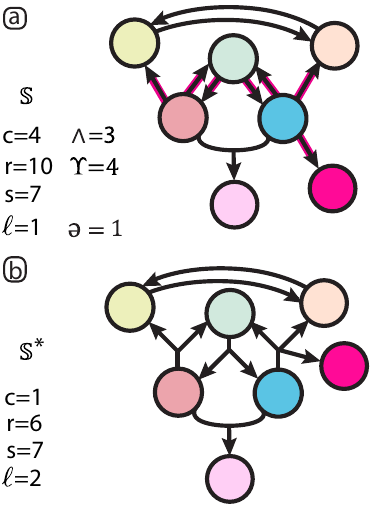}
\caption{Illustration of the general merging procedure. a) Representation with each reaction represented once ($\mathbb{S}$). Reactions with pairwise proportional partners are highlighted in fuchsia. b) Representation after merging pairwise proportional reactions ($\mathbb{S}_{\ffor}$). $\copro = 4$ reactions were merged with respect to $\mathbb{S}$. This results in one emanant and three broken cycles.}
\label{fig:copro_derivation}
\end{figure}

We write all reactions as individual irreversible reactions (splitting up reversible reactions into irreversible pairs) and merge all reactions with proportional rates, so that no pair of currents that remains is pairwise proportional (Fig. \ref{fig:copro_derivation}a-c). We denote $\copro$ the number of pairwise mergers needed to this end and $\mathbb{S}^{*}$ as the resulting stoichiometric matrix. In the derivation of most kinetic schemes (mass action, Michaelis-Menten, Lindemann-Haldane \ref{appendix:co-production beyond mass action}) pairwise proportional rates directly follow from having the same reactant stoichiometry, i.e. the same reactant complex. In matrix form, these reactions have identical columns in $\mathbb{S}^{\ominus}$.

More formally, we use $\copro$ to denote the number of pairwise mergers of reactions. Starting from $\mathbb{S}$, we first partition all $r$ reactions in $r-\copro$ disjoint sets $P^{(q)}$, with the condition that reactions $r_i,r_j$ in the same set have pairwise proportional rates, and pairwise proportional reactions are always in the same single set, i.e., $J_i \propto J_j \implies \exists q \ \ s.t. \ \ i,j \in P^{(q)}, i,j \notin P^{(k)}, (k \neq q)$.

For each set $P^{(q)}$, we can now introduce a single total current $J_{(q)}$, and express the current of each underlying reaction as a fraction of rate constants:

\bea
J_{(q)} &=& \sum_{i \in P^{(q)}} J_i \\
J_i &=& p_i J_{(q)}  \\
p_i &=&  \frac{\kappa_i }{ \sum_{j \in P^{(q)}} \kappa_j } . 
\eea

We can, by the same token, now express the reaction stoichiometry as a single reaction, through the weighted sum:
\bea
\sum_{i \in P^{(q)}} p_i \sum_{k=1}^s  \mathbb{S}_{k,i}^{\ominus} \ce{X}_k \rightarrow \nonumber \\
\ \ \ \ \ \ \ \ \ \ \ \ \sum_{i \in P^{(q)}} p_i
\sum_{k=1}^s  \mathbb{S}_{k,i}^{\oplus} \ce{X}_k . \label{equation:reactionq} \\
\rotatebox{90}{$\equiv$} \ \ \ \ \ \ \ \ \ \ \ \ \ \ \ \ \nonumber\\
\sum_{k=1}^s  \mathbb{S}_{k,q}^{*\ominus} \ce{X}_k \rightarrow \sum_{k=1}^s  \mathbb{S}_{k,q}^{* \oplus} \ce{X}_k . \label{equation:reactionq2} 
\eea
The resulting stoichiometric coefficients make up the stoichiometry of a single merged reaction.

We now define $\mathbb{S}^{*}$ as the matrix obtained by this process, and Eq. \eqref{equation:reactionq} provides the coefficients in the q\textsuperscript{th} column of $\mathbb{S}^{*}$.

We recall that from this procedure we can derive the coproduction law. Denoting $\rotatebox[origin=c]{180}{e} = \Delta \ell$, $\wedge = -\Delta c $, we have
\be
\copro = \rotatebox[origin=c]{180}{e} + \wedge,
\ee
$\rotatebox[origin=c]{180}{e}$ $\#$ co-production emanants (emergent conservation laws) \\
$\wedge$: $\#$ broken cycles

\section{A hidden conserved quantity in an atmospheric chemistry model}
\label{section:atmosphericexample}

As a final application example, we revisit the atmospheric model considered in  recent studies\cite{sturm_conservation_2022,Liu_PRE_2024}, in which a non-integer quantity that appeared to be conserved was discovered numerically. It was unclear whether this was an approximate conservation law, or a genuine one, and no clear interpretation could be given for its non-integer nature. We address both questions below by showing that it is an instance of co-production conservation.

The model starts by considering the irreversible reactions
\bea
\ce{NO2} &+& h\nu \rightarrow \ce{NO} + \ce{O} \\
\ce{O} &+& \ce{O2} \rightarrow \ce{O3}  \\
\ce{O3} &+& \ce{NO} \rightarrow \ce{NO2} + \ce{O2}  \\
\ce{HCHO} &+& 2 \ce{O2} + h\nu \rightarrow 2 \ce{HO}^\bullet_2 + \ce{CO}  \\
\ce{HCHO} &+& h\nu \rightarrow \ce{H2} + \ce{CO} \\
\ce{HCHO} &+& \ce{HO}^\bullet \rightarrow \ce{HO}^\bullet_2 + \ce{CO} + \ce{H2O} \\
\ce{HO}^\bullet_2 &+& \ce{NO} \rightarrow \ce{HO}^\bullet + \ce{NO2} \\
\ce{HO}^\bullet &+& \ce{NO2} \rightarrow \ce{HNO3} \\
\ce{HO2H} &+& h\nu \rightarrow 2 \ce{HO}^\bullet \\
\ce{HO2H} &+& \ce{HO}^\bullet \rightarrow \ce{H2O} + \ce{HO}^\bullet_2 
\eea
It is further supposed that $\ce{H_2O}$ is not monitored, and that $\ce{O2}$ is a reservoir species (chemostatted). The subnetwork thus afforded is 
\bea
\ce{NO2} + h\nu &\rightarrow& \ce{NO} + \ce{O} \\
\ce{O} &\rightarrow& \ce{O3}  \\
\ce{O3} + \ce{NO} &\rightarrow& \ce{NO2}   \\
\ce{HCHO} + h\nu &\rightarrow& 2 \ce{HO}^\bullet_2 + \ce{CO}  \\
\ce{HCHO} + h\nu &\rightarrow& \ce{H2} + \ce{CO} \\
\ce{HCHO} + \ce{HO}^\bullet &\rightarrow& \ce{HO}^\bullet_2 + \ce{CO} \\
\ce{HO}^\bullet_2 + \ce{NO} &\rightarrow& \ce{HO}^\bullet + \ce{NO2} \\
\ce{HO}^\bullet + \ce{NO2} &\rightarrow& \ce{HNO3} \\
\ce{HO2H} + h\nu &\rightarrow& 2 \ce{HO}^\bullet \\
\ce{HO2H} + \ce{HO}^\bullet &\rightarrow&  \ce{HO}^\bullet_2 
\eea
Since $\ce{H2O}$ only occurs as a sink species, the effects of concealing it and chemostatting it are equivalent. %($b_{\ce{X}}=1$ or $b_\square=1$). 

\begin{figure}[tbhp!]
\centering
\includegraphics[width=1.0\linewidth]{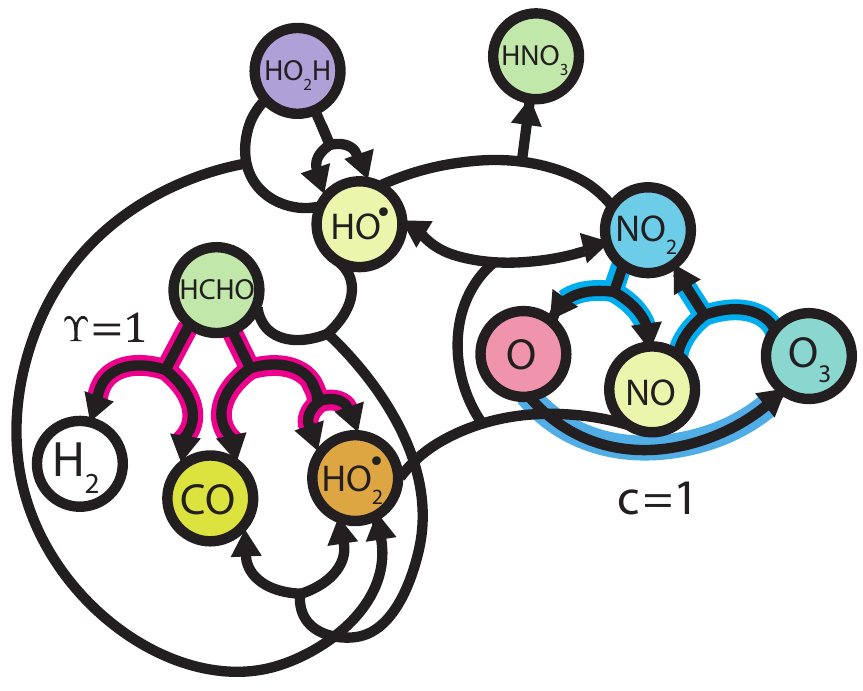}
\caption{Figure 1 from the main text, reproduced for convenience. An Atmospheric CRN model with co-production index $\copro = 1$. Collinear reactions are highlighted in pink. These do not include cycle reactions -  highlighted in blue - and, hence, no cycles break upon merging collinear reactions $\wedge = 0$.  $\rotatebox[origin=c]{180}{e} = 1$  co-production conservation law then results from the co-production law $\copro= \rotatebox[origin=c]{180}{e} + \wedge$, confirming the newly found law $\text{CQ}_3$ marks the detection of a genuine conservation law. }
\label{fig:atmosphericnetworkb}
\end{figure}

We will now see how coproduction emerges: we remove $\ce{O}_2$ and $\ce{H_2O}$ from the description by chemostatting. By the chemostatting of $\ce{O}_2$, reactions $r_4$ and $r_5$ have become collinear, hence $\copro = 1$. The chemostat law\cite{polettini_irreversible_2014}, chemostats = broken conservation laws + affinities reads
\be
s^Y = b + a
\ee
applied here says $s^{\ce{Y}}=2$ reservoirs are introduced, and here one can readily check that $b=2$ stoichiometric conservation laws are broken (i.e., hydrogen and oxygen conservation being lost). Since $\rotatebox[origin=c]{180}{e} = 1$, the number of conservation laws is only reduced by 1 rather than 2. The subnetwork with highlighted collinear reactions is depicted in Fig. \ref{fig:atmosphericnetworkb}.

With stoichiometric matrix
\hspace*{-0.75cm}\vbox{\be
\mathbb{S} = \left(\begin{smallmatrix} 0 & 1 & -1 & 0 & 0 & 0 & 0 & 0 & 0 & 0 \\
1 & 0 & -1 & 0 & 0 & 0 & -1 & 0 & 0 & 0  \\
-1 & 0 & 1 & 0 & 0 & 0 & 1 & -1 & 0 & 0 \\
0 & 0 & 0 & \colorbox{pink}{-1} & \colorbox{pink}{-1} & -1 & 0 & 0 & 0 & 0 \\
0 & 0 & 0 & \colorbox{pink}{2} & 0 & 1 & -1 & 0  & 0 & 1 \\
0 & 0 & 0 & 0 & 0 & 0 & 0 & 0 & -1 & -1 \\
0 & 0 & 0 & 0 & 0 & -1 & 1 & -1 & 2 & -1 \\
1 & -1 & 0 & 0 & 0 & 0 & 0 & 0 & 0 & 0 \\
0 & 0 & 0 & 0 & 0 & 0 & 0 & 1 & 0 & 0 \\
0 & 0 & 0 & \colorbox{pink}{1} & \colorbox{pink}{1} & 1 & 0 & 0 & 0 & 0 \\
0 & 0 & 0 & 0 & \colorbox{pink}{1} & 0 & 0 & 0 & 0 & 0
\end{smallmatrix}\right)   \ \ \ \textcolor{teal}{\begin{smallmatrix}  \ce{O3} \\  \ce{NO} \\ \ce{NO2} \\ \ce{HCHO} \\ \ce{HO2} \\ \ce{HO2H} \\ \ce{OH} \\ \ce{O} \\ \ce{HNO3} \\ \ce{CO} \\ \ce{H2}   \end{smallmatrix}} \ \ .
\ee}
Which has $\text{rk} \left( \mathbb{S}_\triangleright \right) = 9$. Since $r_{\triangleright}=10$, there is $c_\circ=1$ cycle, involving $r_1$ to $r_3$ 
\be
\pmb{c} = (1,1,1,0,0,0,0,0,0,0)
\ee
which is highlighted in Fig. \ref{fig:atmosphericnetworkb}. Reactions $r_4,r_5$ have collinear rates
\be
J_4 = \kappa_4 [\ce{HCHO}] \propto J_5 = \kappa_5 [\ce{HCHO}]
\ee
and we can thus merge them 
\be
J_* = p J_4 + (1-p) J_5 , \ \ \ p = \frac{\kappa_4}{\kappa_4 + \kappa_5}
\ee
which becomes
\be
\ce{HCHO} + hv \rightarrow 2 p \ce{HO2}^\bullet + \ce{CO} + \left(1-p\right) \ce{H2}
\ee
So that our stoichiometric matrix $\mathbb{S}_{*}$ in terms of independent reactions becomes
\hspace*{-0.75cm}\vbox{\be
\mathbb{S}_{*} = \left(\begin{smallmatrix} 
0 & 1 & -1 & 0 & 0 & 0 & 0 & 0 & 0 \\
1 & 0 & -1 & 0 & 0 & -1 & 0 & 0 & 0  \\
-1 & 0 & 1 & 0 & 0 & 1 & -1 & 0 & 0 \\
0 & 0 & 0 & -1 & -1 & 0 & 0 & 0 & 0 \\
0 & 0 & 0 & 2 p & 1 & -1 & 0  & 0 & 1 \\
0 & 0 & 0 & 0  & 0 & 0 & 0 & -1 & -1 \\
0 & 0 & 0 & 0  & -1 & 1 & -1 & 2 & -1 \\
1 & -1 & 0 & 0  & 0 & 0 & 0 & 0 & 0 \\
0 & 0 & 0 & 0  & 0 & 0 & 1 & 0 & 0 \\
0 & 0 & 0 & 1  & 1 & 0 & 0 & 0 & 0 \\
0 & 0 & 0 & 1-p & 0 & 0 & 0 & 0 & 0 \end{smallmatrix}\right). \nonumber
\ee}
As we have not merged any cycle reactions, no cycles are lost $\wedge = 0$. This can readily be checked by making the cycle vector $\pmb{c}$ one reaction shorter 
\bea
\pmb{c} &=& (1,1,1,0,0,0,0,0,0)^T, \\
\mathbb{S}_{*} \pmb{c} &=& \pmb{0}.
\eea
Co-production $\text{\footnotesize{$\copro$}} = \rotatebox[origin=c]{180}{e} + \wedge = 1$ then yields $\rotatebox[origin=c]{180}{e}=1$, and thus an emergent left nullvector
\hspace*{-0.75cm}\vbox{\be
\pmb{\ell}_{*}^{(3)} = \left(6,-5,1,3,9,6,3,6,4,-3,\frac{6 - 18 p}{1-p} \right)^T. \ \ \ \
\ee}
To see if this is the elusive hidden conservation law\footnote{$\text{CQ}_3$ was taken from the Supp. Matt. of Ref\cite{Liu_PRE_2024} where it is derived. It differs from the expression for $\text{CQ}_3$ due to a small typo for the coefficient of [$\ce{OH}^\bullet$]} $\text{CQ}_3$ reported in the literature\cite{Liu_PRE_2024}, we substitute ($p=0.40541.. $):
\hspace*{-0.75cm}\vbox{\bea
\pmb{\ell}_{*}^{(3)} &=& \left(6,-5,1,3,9,6,3,6,4,-3, 2.18.. \right)^T \nonumber \\
CQ_3 &\approx& \left(6,-5,1,3,9,6,3,6,4,-3, 2.21 \right)^T \nonumber
\eea}
The numerical estimate of the measured quantity $\text{CQ}_3$ thus closely approximates the genuine conservation law. We can thus confirm that the observation of $\text{CQ}_3$ was due to a genuine conserved quantity in the (chemostatted) model. The existence and exact nature - co-production conservation - of this conservation law has thereby been clarified. \\

In general, the SID algorithm proposed in that work\cite{Liu_PRE_2024} detects conservation laws of first order differential equations $d_t \pmb{x} = \pmb{f}(\pmb{x})$ by creating a list of $K$ independent phase-space functions  $\pmb{b} = b^{(1)}(\pmb{x}),...,b^{(K)}(\pmb{x})$. $H(\pmb{x})= \pmb{\theta}^T b(\pmb{x})$  is then a conserved quantity, if, for all
$\pmb{x}$:\footnote{As SID is a numerical algorithm, this is done in practice by checking $P$ randomly chosen points. As long as the phase space functions and the flow equations are polynomial, as is the case here, sufficiently large $P$ will ensure that $H(x)$ is conserved for all $\pmb{x}$} 
\bea
g(\pmb{x})^T \pmb{\theta} &=& 0, \\
g(\pmb{x}) &=& \nabla \pmb{b} \pmb{f}(\pmb{x}). 
\eea
Applied to CRNs where $\pmb{x} = [\pmb{\ce{X}}]$, $\pmb{f}(\pmb{x}) = \mathbb{S} \pmb{J} $ and with a list of candidate functions $\pmb{b} = [\pmb{\ce{X}}]$, the above equation reduces to (the transpose of) the condition $\pmb{\ell}^T \mathbb{S} \pmb{J} = 0$ mentioned before. We conclude that the analysis provided herein is sufficient to explain all true linear conservation laws the SID algorithm might find in a CRN. There can be further conservation laws due to special choices of parameters (Sec. \ref{appendix:parametric_collinearity}).  

\section{Markov Processes and co-production}
\label{appendix:markov-processes}

A Markov process (with countable state space) can always be mapped to a CRN by mapping states to species. Ipso facto, our results, such as the co-production law - naturally apply to Markov processes. We will illustrate some applications of co-production below.

\subsection{1d Biased diffusion}
Let us consider a biased random walk on a 1d lattice with n sites, with site 0 and site $n-1$ being absorbing states. 
When jumping away from a site $[k]$, we jump to $[k-1]$ with probability $p$, and to site $[k+1]$ with probability $(1-p)$.
We can represent the two possible jumps to neighboring sites as irreversible reactions, which we can merge
\bea
\ [k-1] &\leftarrow& [k] \rightarrow [k+1], \ \ \ (1 \leq k \leq n-1) \nonumber \\ 
\ [k] &\rightarrow& p  [k-1] + (1-p) [k+1], \ \ 
\eea
There are $n-2$ sites where we can jump away from, and we can thus perform $n-2$ such mergers.
A system with $n$ sites ('species'), thus has $n-2$ independent reactions ($c=0$) and ipso facto there are $\ell = 2$ conservation laws. A stoichiometric matrix for $n=7$ looks as follows
\be
\mathbb{S}^* = \begin{pmatrix} 
p   & 0    & 0   & 0   & 0 \\
-1  & p    & 0   & 0   & 0 \\
1-p & -1   & p   & 0   & 0 \\
0   & 1-p  & -1  & p   & 0 \\
0   & 0    & 1-p & -1  & p \\
0   & 0    & 0   & 1-p & -1 \\
0   & 0    & 0   & 0   & 1-p 
\end{pmatrix}
\ee
Conservation of probability directly yields $\pmb{\ell}^{(1)} = (1,1,1,1,1,1,1)$ as left nullvector and conserved quantity
\be
\pmb{\ell}^{(1)} \cdot \pmb{p} = \sum_k p_k (t) = 1
\ee
For the second conserved quantity $\pmb{\ell}^{(2)} = (b_0, b_1, .., b_{n-1})$, we can freely set $b_0 = 0, b_1=1$ by scaling and subtraction of $\pmb{\ell}^{(1)}$. Then from inspection of $\pmb{\ell}^{(2)} \mathbb{S} = \pmb{0}$
\bea
 - 1 + (1-p) b_2 = 0, \\
p b_{k-1} - b_k + (1-p) b_{k+1} = 0, \\
b_k = \frac{1-p}{p} b_{k-1} + 1
\eea
from which we obtain
\bea
q &=& \frac{1-p}{p}, \\
b_k &=& q^{k - 1} + \frac{q^k - 1}{q - 1}
\eea
At long times, only site $0$ and site $k-1$ will be populated. Let us denote with an overbar $\bar{p}_{k}$
\be
\bar{p}_{k} = \lim_{t \rightarrow \infty} p_k (t)
\ee
the conservation law $\pmb{\ell}^{(2)}$ informs of a constraint on the dynamics of the random walk, and in the long-time limit provides the stationary occupations
\bea
L^{(2)} &=& \pmb{\ell}^{(2)} \cdot \pmb{p}(0) = \pmb{\ell}^{(2)} \cdot \pmb{p}(t) \\
&=& \pmb{\ell}^{(2)} \cdot \bar{\pmb{p}} = b_{n-1} \bar{p}_{n-1}. \nonumber
\eea
For illustration, let $p=0.2$. We then have
\be
\pmb{\ell}^{(2)} = \left(0, 1, \frac{5}{4}, \frac{21}{16}, \frac{85}{64}, \frac{341}{256}, \frac{1365}{1024} \right)
\ee
We initially occupy site $[2]$ and $[3]$ with equal probability $p_2(0)=p_3(0)=1/2$, so that 
\be
L^{(2)} = \frac{5}{4} p_2(0) + \frac{21}{16} p_3(0) = \frac{41}{32}
\ee
For $n=7$ sites, our stationary probability for reaching absorbing site $[6]$ is now 
\bea
\bar{p}_6 = L^{(2)} / b_6 &=& \frac{1024}{1365} \frac{41}{32} \\
&=& \frac{1312}{1365} \approx 0.961 \nonumber
\eea

\subsection{2d diffusion}
This logic applies more universally. Consider some connected lattice on which diffusion is taking place, with $n$ distinct sites, among which $k \geq 1$ are absorbing states. There are then $n-k$ nonabsorbing sites, with ipso facto outgoing reactions which can be merged to a single reaction, so that there are $n-k$ independent ($c=0$) reactions. From SL1, there are then $\ell = k$ conservation laws, i.e. there are as many conservation laws as there are sinks. Furthermore, we can always choose for our conservation laws a basis such that each has a single nonzero coefficient corresponding to a sink.

Let us now consider a 2d example, namely a 3-by-3 lattic with periodic boundary conditions. We thus have reactions of the form
\bea
\ [x,y] &\rightarrow& [x \pm 1,y], \ \  [x,y] \rightarrow [x, y \pm 1] \nonumber \\
\ [x,y] &\rightarrow& \frac{1}{4} [x + 1,y] + \frac{1}{4} [x - 1,y] \nonumber \\
&+& \frac{1}{4} [x,y + 1] + \frac{1}{4} [x,y - 1] 
\eea
where coordinates are evaluated modulo 3. Let us now make $[1,0]$ and $[2,2]$ absorbing sites. We then have
\be
\! \! \! \! \! \! \! \! \! \! \!  \mathbb{S}^{*} = \begin{pmatrix}
          -1  & \frac{1}{4} & \frac{1}{4} & 
           0 & 0 & \frac{1}{4} & 0  \\  
           
          \frac{1}{4} & \frac{1}{4} & 0 & 
           \frac{1}{4} & 0 & 0 & \frac{1}{4} \\
           
          \frac{1}{4} & -1 & 0 & 
           0 & \frac{1}{4} & 0 & 0 \\  
           
          \frac{1}{4} & 0 & -1 & \frac{1}{4} & 
           \frac{1}{4} & \frac{1}{4} & 0 \\ 
           
          0 & 0 & \frac{1}{4} &  -1 & 
           \frac{1}{4} & 0 & \frac{1}{4} \\ 
           
          0 & \frac{1}{4} & \frac{1}{4} & \frac{1}{4} & 
           -1 & 0 & 0 \\ 
           
          \frac{1}{4} & 0 & \frac{1}{4} & 0 & 0 & -1 & \frac{1}{4} \\

          0 & 0 & 0 & \frac{1}{4} & 0 & \frac{1}{4} & -1 \\
          
          0 & \frac{1}{4} & 0 & 0 & \frac{1}{4} & \frac{1}{4} & \frac{1}{4} 
          
      \end{pmatrix}
    \begin{matrix}
    [0,0] \\
    [1,0] \\
    [2,0] \\
    [0,1] \\
    [1,1] \\
    [2,2] \\
    [0,2] \\
    [1,2] \\
    [2,2]     
    \end{matrix}
\ee
One convenient basis for the left nullspace is
\bea
\pmb{\ell}^{(1)} = (6,10,5,5,6,4,4,5,0)\\ %(1,1,1,1,1,1,1,1,1) \\
\pmb{\ell}^{(2)} = (4,0,5,5,4,6,6,5,10) 
\eea
from $\pmb{\ell}^{(1)}$ it can immediately be seen that the stationary probability to reach absorbing site $[1,0]$ $(\bar{p}_{1,0})$ from a single starting site can be $0.4$, $0.5$ or $0.6$. From $\pmb{\ell}^{(2)}$ we draw the same conclusion for $\bar{p}_{2,2}$, and we can furthermore see that $\bar{p}_{2,2} = 1 - \bar{p}_{1,0}$.

\subsection{n-mer adsorption}

We consider the irreversible adsorption of a n-mers on an initially empty lattice of size L. We can now denote $n_k$ the population of gaps of k empty sites between adsorbed n-mers and/or lattice extremities, so that initially, $p(n_L=1,t=0) = 1$.
In terms of gap populations, the adsorption process resembles a fragmentation process:
\be
\ [m+n] \rightarrow [m-k] + [k] \ \ \ (0 \leq k \leq m+1)
\ee
merging collinear reactions, we have
\bea
\ [m+n] &\rightarrow& \frac{1}{m+1} \sum_{k=0}^{m+1} \left( [m-k] + [k] \right) \nonumber \\
\ [m+n] &\rightarrow& \frac{2}{m+1} \sum_{k=0}^{m+1} [k] 
\eea
which in a stoichiometric matrix yield columns of the form $(-1, [0]_n, [\frac{2}{m+1}]_{m+1}$), for instance, for $L=10$, $n=3$ we have
\be
\mathbb{S}^{*} = \begin{pmatrix}
\frac{2}{7} & \frac{2}{6} & \frac{2}{5} & \frac{2}{4} & \frac{2}{3} & 1 & 2 \\

\frac{2}{7} & \frac{2}{6} & \frac{2}{5} & \frac{2}{4} & \frac{2}{3} & 1 & 0  \\		  

\frac{2}{7} & \frac{2}{6} & \frac{2}{5} & \frac{2}{4} & \frac{2}{3} & 0 & 0  \\

\frac{2}{7} & \frac{2}{6} & \frac{2}{5} & \frac{2}{4} & 0 & 0 & -1 \\
          
\frac{2}{7} & \frac{2}{6} & \frac{2}{5} & 0 & 0 & -1 & 0 \\ 

\frac{2}{7} & \frac{2}{6} & 0 & 0 & -1 & 0 & 0 \\

\frac{2}{7} & 0 & 0 & -1 & 0 & 0 & 0 \\

0  & 0 & -1 & 0 & 0 & 0 & 0  \\  

0  & -1 & 0 & 0 & 0 & 0 & 0  \\  

-1  & 0 & 0 & 0 & 0 & 0 & 0  
      \end{pmatrix}   \ \ \ \textcolor{teal}{\begin{matrix} [0] \\ [1] \\ [2] \\ [3] \\ [4] \\ [5] \\ [6] \\ [7] \\ [8] \\ [9] \end{matrix}} \ .  
\ee
Gaps of size smaller than $n$ can not be further reduced, whereas gaps of size $n$ or greater always have an outgoing reaction. We thus have $n$ coproduction conservation laws $\pmb{\ell}^{(0)},..,\pmb{\ell}^{(n-1)}$ governing state probabilities (see Eq. \eqref{equation:gensol_nmerb} for a general solution),  satisfying $\pmb{\ell}^T  \mathbb{S} = \pmb{0}^T$. A convenient basis for these conservation laws is one where each vector has a nonzero argument for only a single of the irreducible gaps:

\bea
\pmb{\ell}^{(0)} &=& \left( 1, 0, 0, 2, 1, \frac{2}{3}, \frac{3}{2}, \frac{8}{5}, \frac{14}{9}, \frac{37}{21} \right)^T ,
%\left(\frac{37}{21}, \frac{14}{9}, \frac{8}{5},\frac{3}{2}, \frac{2}{3}, 1, 2, 0, 0, 1 \right)^T,
\nonumber \\
\pmb{\ell}^{(1)} &=& \left( 0, 1, 0, 0, 1, \frac{6}{9}, \frac{1}{2}, \frac{4}{5}, \frac{8}{9}, \frac{19}{21} \right)^T ,
%\left(\frac{19}{21}, \frac{8}{9}, \frac{4}{5}, \frac{1}{2}, \frac{6}{9}, 1, 0, 0, 1, 0 \right)^T,
\nonumber \\ 
\pmb{\ell}^{(2)} &=&  \left(0,0,1,0,0,\frac{6}{9},\frac{1}{2},\frac{2}{5},\frac{5}{9},\frac{13}{21} \right)^T. 
%\left(\frac{13}{21}, \frac{5}{9}, \frac{2}{5}, \frac{1}{2}, \frac{6}{9}, 0, 0, 1 , 0, 0 \right)^T
\label{equation:coeffcons_law_nmerb}
\eea
The value of conserved quantity  $L^{(1)},L^{(2)},L^{(3)}$ follows from evaluation of the initial condition
\be
L^{(k)} = \pmb{\ell}^{(k)} \cdot \begin{pmatrix} \langle [0]_t \rangle \\
\langle [1]_t \rangle \\
\langle [2]_t \rangle \\
\vdots \\
\langle [L-1]_t \rangle \\
\langle [L]_t \rangle  \end{pmatrix} . \label{equation:cons_avg_nmerb}
\ee

\begin{figure}[tbhp!]
\centering
\includegraphics[width=1.0\linewidth]{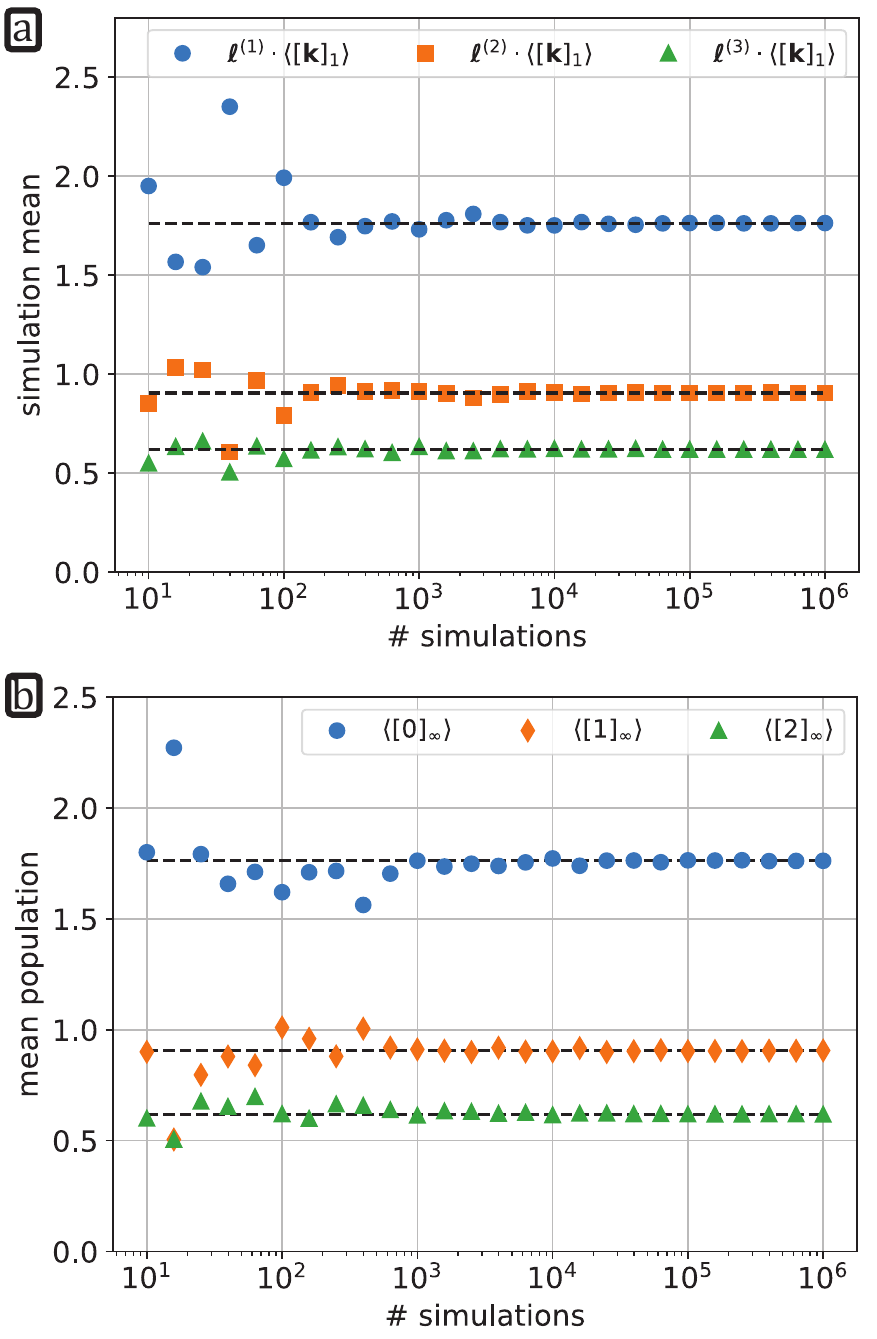}
\caption{a) Evaluation of conserved quantities over averages after 1 reaction, by averaging over an increasing number of simulations. As the number of simulations increases, convergence occurs to the theoretical values (dotted line) of $L^{(1)}, L^{(2)}, L^{(3)}$. b) Averaged number of gaps $[0],[1],[2]$ as a function of the number of individual simulated realizations, for trimer ($n=3$) adsorption on a lattice of size $L=9$. Dotted lines correspond to theoretical expectation values $L^{(1)}$, $L^{(2)}$, $L^{(3)}$, to which the simulated averages converge. Note that the conserved quantities here behave as regular conservation laws on the level of deterministic ODEs. Co-production conservation laws do not constrain individual realizations of a stochastic process, but do constrain the ensemble of realizations as a linear law on expectation values (Eq. \eqref{equation:cons_avg_nmerb}).}
\label{fig:n-merplot}
\end{figure}

The conserved quantity pertains to averages and not individual trajectories. This is illustrated in Fig. \ref{fig:n-merplot}a, showing an evaluation of the conserved quantity after $t=1$ reaction, averaging over an increasing number of simulations. Convergence to the $L^{(1)},L^{(2)},L^{(3)}$ occurs as the number of simulations is increased to approach the theoretical expectation value.

Initially starting with an empty lattice of size $L=9$, ($[k](t=0)=\delta_k^L$, i.e., $[9]_0=1, [k]_0=0 \ (k<9)$), at long times $t\rightarrow \infty$, we then attain as stationary expectation value for the gap population $\langle [0]_\infty \rangle$

\bea
\langle [0]_\infty \rangle &=& \langle [\pmb{k}]_\infty \rangle \cdot \pmb{\ell}^{(1)}  =  L^{(1)}_{t \rightarrow \infty} = L^{(1)}_{t=0} \nonumber \\
&=&  \langle [\pmb{k}]_0 \rangle \cdot \pmb{\ell}^{(1)} = \langle [9]_0\rangle \ell^{(1)}_{10} = L^{(1)} \nonumber \\
&=&  \frac{37}{21} 
\eea
The same argument for $\langle [1]_\infty \rangle, \langle [2]_\infty \rangle$ results in
\bea
\langle [1]_\infty \rangle &=& \langle [9]_0\rangle \ell^{(2)}_{10} = L^{(2)} =  \frac{19}{21} \\
\langle [2]_\infty \rangle &=&  \langle [9]_0\rangle \ell^{(3)}_{10} = L^{(3)} = \frac{13}{21} , 
\eea
where $\langle [k]_\infty \rangle$ denotes the expectation value for the number of gaps of size $k$ for $t \rightarrow \infty$. Simulations of the n-mer model for $L=9, n=3$ are consistent with this result, and simulated mean gap size populations are shown in Fig. \ref{fig:n-merplot}b. The jamming coverage $\rho_{n,L}$ can then be written as
\be
\rho_{n,L} = 1 - \sum_{k=1}^{n} \frac{k L^{(k)}}{L}
\ee
For our example, we obtain $\rho_{3,9} = \frac{144}{189}\approx 0.76190$, which is noticably smaller than the $L\rightarrow \infty$ limit $\rho_{3,\infty}=0.82365$

\subsection{General solution for arbitrary \textit{L,n}}

The coefficients of a co-production $\pmb{\ell}^{(q)}$ conservation law must be such that $\pmb{\ell}$ is a left null vector $\mathbb{S}^*$. For the adsorption of n-mers, we then have

\be
\ell^{(q)}_{j+n} = \frac{2}{j} \sum_{k=1}^{j} \ell^{(q)}_{k} .
\ee
Rather than summing over $k=1$ to $k=j$, we can confine our sum from $k=j-n+1$ to $k=j$,

\hspace*{-0.75cm}\vbox{\bea
\ell^{(q)}_{j+n} &=& \frac{2}{j} \left(\sum_{k=j-n+1}^{j} \ell^{(q)}_{k} +  \sum_{k=1}^{j-n} \ell^{(q)}_{k} \right) \nonumber \\
&=& \frac{2}{j} \sum_{k=j-n+1}^{j} \ell^{(q)}_{k} + \frac{j - n }{j} \ell^{(q)}_{j} . \nonumber 
\eea}
We may similarly express $\ell^{(q)}_{j+n-p} $, $0<p<n$ in terms of $\ell^{(q)}_{j-n+1}$ to $\ell^{(q)}_{j}$:

\hspace*{-1.0cm}\vbox{\bea
\ell^{(q)}_{j+n-p} &=& \frac{2}{j-p} \left(\sum_{k=j-n+1}^{j-p} \ell^{(q)}_{k} +  \sum_{k=1}^{j-n} \ell^{(q)}_{k} \right) \nonumber \\
&=& \frac{2}{j-p} \sum_{k=j-n+1}^{j-p} \ell^{(q)}_{k} + \frac{j - n}{j-p} \ell^{(q)}_{j} \nonumber 
\eea}

From these expression, we can then calculate any coefficient in a conservation law for arbitrary oligomer size $n$ and lattice size $L$:
\hspace*{-1.0cm}\vbox{\bea
&\begin{pmatrix} \ell^{(q)}_{mn-n+1} \\ \ell^{(q)}_{mn-n+2} \\ \vdots \\ \ell^{(q)}_{mn-2} \\ \ell^{(q)}_{mn-1}  \\ \ell^{(q)}_{mn}  \end{pmatrix} = \prod_{p=1}^{m-1} \mathbb{F}_{p,n} \begin{pmatrix}\ell^{(q)}_1 \\ \ell^{(q)}_{2} \\ \ell^{(q)}_{3} \\ \vdots \\ \ell^{(q)}_{n-1}  \\ \ell^{(q)}_{n} \end{pmatrix} \label{equation:gensol_nmer} \\ 
&\mathbb{F}_{p,n} = \left(\begin{smallmatrix}
\frac{2}{np-n+1} & 0 & 0  & \dots & 0 &    \frac{np-n}{np-n+1} \\
 \frac{2}{np-n+2} & \frac{2}{np-n+2} & 0 &  \dots & 0 & \frac{np-n}{np-n+2} \\
\vdots & \vdots & \vdots & \ddots & \vdots & \vdots \\
 \frac{2}{np-2} & \frac{2}{np-2} & \frac{2}{np-2}  & \dots & 0 & \frac{np-n}{np-2} \\
 \frac{2}{np-1}  & \frac{2}{np-1} & \frac{2}{np-1} & \dots & \frac{2}{np-1}  & \frac{np-n}{np-1} \\
 \frac{2}{np} & \frac{2}{np} &  \frac{2}{np}  & \dots  & \frac{2}{np} & \frac{np-n+2}{np} 
\end{smallmatrix} \right) . \nonumber
  \eea}
  
For oligomers of size $n$, a convenient base for nullvectors in Eq. \eqref{equation:gensol_nmer} follows from equating the first $n$ coefficients to a unit vector $\hat{\pmb{e}}_q$ of length $n$ for $\ell^{(1)}$ to $\ell^{(m)}$ and we get 

\hspace*{-0.5cm}\vbox{\be 
\begin{pmatrix} \ell^{(q)}_1 & \ell^{(q)}_{2} & \dots & \ell^{(q)}_{n} \end{pmatrix}^T = \hat{\pmb{e}}_q , \ \ \ (0 \leq k \leq n) , 
\ee}
which was the approach also adopted in the main text. With this choice, one can then obtain the stationary expectation values for gap populations for arbitrary $L,n$ as
\be
\langle [k]_\infty \rangle = \langle [L]_0 \rangle \ell_{L+1}^{(k+1)} = L^{(k+1)} .
\ee

\section{Parametric collinearity}
\label{appendix:parametric_collinearity}
We have considered collinearity that is due to structure alone. For certain combinations of parameters and initial conditions, a CRN may become symmetric and behave like a lower-dimensional CRN. As a concrete example, we 
consider
\be
\ce{C} \underset{3}{\overset{1}{\leftrightarrows}} \ce{A} + \ce{B} \underset{4}{\overset{2}{\rightleftarrows}} \ce{D} .
\ee
Which we can ostensibly write in terms of two  reversible currents
\bea
J_{1,\circ} &=& \kappa_1 [\ce{A}] [\ce{B}] - \kappa_3 [\ce{C}] = J_{1} - J_{3} \nonumber \\ 
J_{2,\circ} &=& \kappa_2 [\ce{A}] [\ce{B}] - \kappa_4 [\ce{D}] = J_{2} - J_{4} \nonumber
\eea

and, hence, \textit{a priori}, no collinearity should exist among the reversible currents. 

If we now fix $[\ce{C}]_0 = [\ce{D}]_0$, $\kappa_1=\kappa_3$, $\kappa_2=\kappa_4$, then the rates become collinear $J_{1,\circ} = J_{2,\circ}$ and dynamics for $[\ce{C}]$ and $[\ce{D}]$ become equivalent, hence we have a conserved quantity: 
\be
L = [\ce{C}]-[\ce{D}] = 0.
\ee 
This relates to a symmetry in the CRN $\ce{C} \leftrightarrow \ce{D}$, which vanishes when we let $[\ce{C}]_0 \neq [\ce{D}]_0$. 

In terms of irreversible reactions, we have always have that $J_1 \propto J_2$ always (resulting in $\wedge=1$). An additional parametric collinearity $J_3 \propto J_4$ emerges when $[\ce{C}]_0 = [\ce{D}]_0$.

This requirement of 3 parametric constraints is readily attained in chemical practice: if $\ce{C}$ and $\ce{D}$ are enantiomers in an achiral environment, then the 2 constraints on rate constants will naturally be verified. The constraint $[\ce{C}]_0 = [\ce{D}]_0$ would require that we either add $\ce{C}$ and $\ce{D}$ as a racemic mixture, or do not add them at all. We refer to Ref. \cite{laurent_emergence_2021,Laurent2022} for pertinent further decompositions of CRNs with symmetries due to chiral species. 

Since parametric collinearity occurs for specific choices of rate constants and initial conditions, it will not generically be detected by algorithmic procedures that seek out conserved quantities.

\section{Co-production beyond mass-action}
\label{appendix:co-production beyond mass action}

A key prerequisite for coproduction is that currents can become linearly dependent, which under mass-action naturally allows to relate co-production to reaction stoichiometry. This prerequisite is not satisfied in every setting, for instance parameter-rich kinetics\cite{Vassena2023symbolic} axiomatically excludes such collinearities through additional parameters. 

Nevertheless, coproduction is not restricted to mass action. Moreover, in chemical practice, many non-mass action rate expressions are approximations derived from mass-action expressions. When we perform those same derivations starting from a CRN with co-production, we find that th enumber of co-production emanants $\rotatebox[origin=c]{180}{e}$ is readily retained. As such, co-production is not limited to mass action and can be encountered in many further settings. We provide a few derivations below.  

\subsection{Michaelis-Menten mechanism}

Consider the following enzymaticatically catalyzed co-production scheme, involving two reactions described as irreversible
\bea
\! \! \! \! \! \! \! \! \ce{X} + \ce{E} \overset{1}{\leftarrow} \ce{EX} \overset{2}{\leftrightarrows} \ce{E} + \ce{S} \overset{3}{\leftrightarrows} \ce{EY} \overset{4}{\rightarrow} \ce{Y} + \ce{E} . 
\eea
we find $r=6, s=6, \ell=2, c= 2$, and  $\copro=3$, Resulting $\rotatebox[origin=c]{180}{e}=1$ emanant and $\wedge=2$ broken cycles.

Now consider we have as our approximation the usual Michealis-Menten setup
\bea
\! \! \! \! \! \! \! \! \! \! \! d_t [\ce{EX}] &=& -\left(\kappa_1 + \kappa_2^-\right) [\ce{EX}] + \kappa_2^+ [\ce{E}][\ce{S}] = 0 \nonumber \\
\! \! \! \! \! \! \! \! \! \! \! d_t [\ce{EY}] &=& -\left(\kappa_4 + \kappa_3^-\right) [\ce{EY}] + \kappa_3^+ [\ce{E}][\ce{S}] = 0 \nonumber \\
\! \! \!  \! \! \! \! \! \! \! \! [\ce{E}]^0 &=& [\ce{E}] \left(1 + \frac{\kappa_2^+ [\ce{S}]}{\kappa_1 + \kappa_2^-} + \frac{\kappa_3^+ [\ce{S}]}{\kappa_4 + \kappa_3^-} \right)
\eea
And we now have as an approximate non mass action description
\bea
\ce{X} &+& \ce{E} \overset{1}{\leftarrow} \ce{E} + \ce{S}  \overset{4}{\rightarrow} \ce{Y} + \ce{E} .  \\
J_1 &=& \kappa_1 [\ce{EX}] = \frac{\kappa_1 [\ce{E}]^0 [\ce{S}]}{\left(1 + \frac{\kappa_2^+ [\ce{S}]}{\kappa_1 + \kappa_2^-} + \frac{\kappa_3^+ [\ce{S}]}{\kappa_4 + \kappa_3^-} \right)} \nonumber \\
J_4 &=& \kappa_4 [\ce{EY}] = \frac{\kappa_4 [\ce{E}]^0 [\ce{S}]}{\left(1 + \frac{\kappa_2^+ [\ce{S}]}{\kappa_1 + \kappa_2^-} + \frac{\kappa_3^+ [\ce{S}]}{\kappa_4 + \kappa_3^-} \right)}  \nonumber 
\eea
Here, we have $\copro=1$  and $\rotatebox[origin=c]{180}{e}=1$. The dimension reduction by coproduction is here preserved upon taking the Michaelis-Menten approximation. 

\subsection{Lindemann-Haldane mechanism}

The Haldane-Lindemann scheme describes the activation and deactivation of a species by a molecular collision (usually in the gas phase) with a species $\ce{M}$, with the activated species undergoing a further unimolecular reaction. A coproduction analogue is then 
\bea
&\ \ce{A} + \ce{M} \overset{1}{\leftrightarrows} \ce{A}^* + \ce{M}, \\
&\ce{X} \overset{2}{\leftarrow} \ce{A}^* \overset{3}{\rightarrow} \ce{Y}. 
\eea
for which $r=4$, $\copro=1$, $\rotatebox[origin=c]{180}{e}=1$. 
\be
L^* = \kappa_3 [\ce{X}] + \kappa_2 [\ce{Y}]
\ee
We may now write
\bea
d_t [\ce{A}^*] &=& \kappa_1^+ [\ce{A}] [\ce{M}] - \kappa_1^- [\ce{A}^*] [\ce{M}] \nonumber \\  &-&  \kappa_2 [\ce{A}^*] - \kappa_3 [\ce{A}^*] = 0 \nonumber \\
\ [\ce{A}^*] &=& \frac{\kappa_1^+ [\ce{A}][\ce{M}] }{\kappa_1^- [\ce{M}] + \kappa_2  + \kappa_3 }
\eea
and this affords effective expressions for the currents
\bea
J_2 &=& \kappa_2 \frac{\kappa_1^+ [\ce{A}][\ce{M}] }{\kappa_1^- [\ce{M}] + \kappa_2  + \kappa_3 } \\
J_3 &=& \kappa_3 \frac{\kappa_1^+ [\ce{A}][\ce{M}] }{\kappa_1^- [\ce{M}] + \kappa_2  + \kappa_3 } 
\eea
which retains $\rotatebox[origin=c]{180}{e}=1$.

It may be noted more generally that these derivations involve making approximations to yield an expression for some quasi-stationary concentration, which is then substituted. For a collinear set of currents, performing these substitutions cannot lift the collinearity. By the same token it is easy to verify that analogous arguments hold to preserve emanants in other commonly encountered rate laws, for instance in the Eigen-Wilkins mechanism (associative substitution) and in dissociative substitution.

\section{From Markov process to stoichiometric matrix}
\label{Appendix:Markov-to-SM}

We follow the setup of discrete Langevin processes as described in 'Stochastic Energetics' by Ken Sekimoto\cite{Sekimoto_stoch_en}. We subsequently consider the case of Markov-Processes satisfying a 'population picture' and derive its corresponding population representation. For a more general formalism and treatment we refer the interested reader to the second-quantization of classical processes \textit{sensu} Doi-Peliti\cite{Peliti1985,Doi1976a,Doi1976b}.

\subsection{Discrete Langevin Equation, Master Equation}
\label{Appendix:DLE}

This section largely follows the derivation and notation of a more detailed discussion in \cite{Sekimoto_stoch_en}. We first consider individual trajectories / stochastic realizations of a process, as one would e.g. obtain in a single-molecule experiment or a simulation with Gillespie's algorithm.  

We start out with Poisson noise 
\be
\zeta(t) = \sum_{\alpha} \delta(t-t_\alpha)
\ee
where $\delta(z)$ a Dirac delta function, so that the occurrence of $n$ successive spike events at is distributed as
\be
P\left[\int_t^{t+\Delta t} \zeta(s) ds = n \right] = \frac{e^{-w\Delta t}}{n!} (w \Delta t)^n
\ee
%for a single realization of poisson noise $\hat{\zeta}$, the realized number of spikes between $t, t+\Delta t$ is then $\hat{n}$  

%and then $\hat{n} \equiv \int_{t}^{t+\Delta t} \hat{\zeta}(s) ds$, and for a Poisson process one has that $\langle \hat{n} \rangle = w \Delta t$. 
The expectation value for the number of spikes in a time interval $\Delta t$ then gives us $\langle n \rangle = w \Delta t$, where $w$ is our jump rate.

The occurrence of transitions between discrete states $|i \rangle, |j\rangle$ (for our purposes: unit vectors) by the Poisson process can be counted by $\hat{n}_{i,j}$ as 
\bea
\hat{n}_{i,j} (t, t+\Delta t) &=& \int_t^{t+\Delta t} \zeta_{i,j}(s) ds \\
\langle \zeta_{i,j}(t) \rangle &=& w_{i \rightarrow j} 
\eea
denoting dual base vectors $\langle i |$, we can write $\langle i | j \rangle = \delta_{i,j}$. Where $\delta_{i,j}$ denotes the Kronecker Delta, $\delta_{j,j} =1, \delta_{i\neq j,j} =1  $

Denoting the state of the system at time $t$ as $|i(t)\rangle$, we can now write the discrete Langevin equation 
\be
d_t | i(t) \rangle = \sum_j \left( |j \rangle - |i(t) \rangle \right) \cdot \zeta_{i(t),j}(t)
\ee
where multiplication $\cdot$ is applied in the It$\hat{\text{o}}$ sense. The difference $\left( |j \rangle - |i \rangle \right)$ denotes a reaction when $w_{i \rightarrow j} \neq 0$. Let us denote $r$ the number of such reactions we have, which we index from $1$ to $r$, i.e. we can assign each rate $w_{i\rightarrow j}$ to a stochastic current $j_{k(i,j)}$, where $k(i,j)$ uniquely maps each reaction to a unique index in $[r]$.  
\be
\! \! \! \! \! \! \! \! \! \forall \zeta_{h,j} > 0, \ \ \  j_{k(h,j)} = \zeta_{h,j} \langle h | i(t) \rangle = \zeta_{h,j} \delta_{h,i(t)}
\ee
where we remark that for a reaction to occur its 'reactant' state must be occupied. 

Let us imagine our initial state at $t=0$ is $|\alpha \rangle$ and a spike arrives at $t^*$ for $\zeta_{\alpha,\beta}$, i.e.
\bea
\ |i(t) \rangle &=& |\alpha \rangle \ \ \ 0 \leq t < t^* \\
\ |i(t^{*+}) \rangle &=& |\beta \rangle 
\eea
this change in state follows from the integration 
\bea
\Delta |i \rangle &=& \int_0^{t^{*-}} d_t | i(t) \rangle = 0  \\
\Delta |i \rangle &=& \int_0^{t^{*+}} d_t | i(t) \rangle = |\beta \rangle - |\alpha \rangle 
\eea

the probability to occupy a state $j$ can be obtained as the average over trajectories 
\bea
\! \! \! P_j (t) &=& \langle \delta_{j,i(t)} \rangle   \\
\! \! \! J_k &=& \langle j_k \rangle = \langle \zeta_{h(k),j} \rangle  \langle \delta_{h(k),i(t)} \rangle \nonumber \\
&=& \kappa_k P_{h(k)} (t)  \nonumber
\eea
and so we have
\be
d_t P_{i} = \sum_{j} \left(P_{j} w_{j \rightarrow i} - P_{i}  w_{ \rightarrow j}\right)
\ee
we now confine our summation to the $r$ reactions. Denoting $h(q)$ (resp. $w(q)$) the state index of the reactant (resp. product) state in reaction $q$,  i.e.
\bea
\ce{X}_{h_{q}} \overset{q}{\rightarrow} \ce{X}_{w_{q}}
\eea
We can now introduce stoichiometric matrices as a bookkeeping device that tells us for the qth transition what its reactant and product state are
\bea
\mathbb{S}_{h(q),q}^{\ominus} = 1, \ \  \mathbb{S}_{j \neq h(q),q}^{\ominus} = 0 \\ 
\mathbb{S}_{w(q),q}^{\oplus} = 1, \ \  \mathbb{S}_{j \neq w(q),q}^{\oplus} = 0 
\eea
and the master equation then becomes
\be
d_t P_{i} = \sum_{q} \left(P_{h(q)} w_{h(q) \rightarrow i} - P_{i}  w_{i \rightarrow w(q)}\right) \nonumber
\ee
where we observe that
\bea
 P_{h(q)} w_{h(q) \rightarrow i} = \pmb{e}_i^T \mathbb{S}^{\oplus} \hat{\pmb{e}}_q \kappa_q P_{h(q)}
\eea
which is nonzero when $i = w(q)$, and 
\bea
 P_{i} w_{i \rightarrow w(q)} &=& \pmb{e}_i^T \mathbb{S}^{\ominus} \hat{\pmb{e}}_q \kappa_q P_{i} \\
 &=& \pmb{e}_i^T \mathbb{S}^{\ominus} \hat{\pmb{e}}_q \kappa_q P_{h(q)}
\eea
which is nonzero when $i = h(q)$. we now introduce a current $J_q$ for reaction $q$
\be
J_q = P_{h(q)} w_{h(Q) \rightarrow w(q)} = \kappa_q P_{h(q)}
\ee
And we can then write
\bea
\pmb{e_i}^T d_t \pmb{P} = \sum_q \pmb{e_i}^T \mathbb{S}^{\oplus} \hat{\pmb{e}}_q J_q - \pmb{e_i}^T \mathbb{S}^{\ominus} \hat{\pmb{e}}_q J_q  \nonumber
\eea
Denoting $\pmb{J} = (J_1, ..., J_r)^T$, we then obtain the system of equations
\bea
d_t \pmb{P} &=& \left(\mathbb{S}^{\oplus} - \mathbb{S}^{\ominus}\right) \pmb{J}  
\eea
which we can rewrite as 
\be
d_t \pmb{P} = \mathbb{S} \pmb{J}(\pmb{P}) = \mathbb{S}^* \pmb{J}^* (\pmb{P}) \label{equation:mastereq}
\ee
where $\mathbb{S}^*, \pmb{J}^*$ correspond to stoichiometric matrix and current vector after mergers as detailed in the main text.
In this picture of singly occupied states, left nullvectors of $\mathbb{S}^*$ correspond to constraints on occupation probabilities $\{P_1, ..., P_s\}$.

By the same derivation, the discrete Langevin equation can be written in terms of a stoichiometric matrix $\mathbb{S}$, which we multiply with a vector of stochastic currents $\pmb{j} = (j_1, ..., j_r)$
\be
d_t | i(t) \rangle = \mathbb{S} \pmb{j} 
\ee
where the matrix-vector product follows the Ito product. Note that performing a 'merger' is nontrivial on the level of single realizations: this will introduce noninteger stoichiometric indices that break down the discrete nature of the discrete Langevin equation. On the level of a single realization with singly occupied states, coproduction counts the number of transitions beyond the first for states undergoing multiple transitions.

\subsection{Population picture}

We continue to consider Markov processes, but now considering states that reflect discrete populations $\{n_i\} = \{n_1, .., n_s\}$ of $s$ species, which undergo transitions that are first order in a single population (except for external influx transitions). A state may now be denoted $\ |n_1 .. n_k .. n_s \rangle$ or simply $| \{n_i\} \rangle$. The discrete Langevin equation now simply follows by summing over all pairs of states
\be
\! \! \! \! \! \! \! \! \! d_t | \{n_i\} \rangle =  \sum_{ \{n_{j}'\} \in N} \left( \{n_{j}'\} \rangle - | \{n_i(t)\} \rangle \right) \cdot \zeta_{n_{i},n_j}(t),
\ee
typically, almost all terms in the sum are zero: a transition exists for only a very small number of these pairs. We can alternatively formulate this in vector form 
\bea
\! \! \! \! \! \! \! \! \! d_t | \pmb{n} \rangle &=&  \sum_{ \pmb{n}' \in \pmb{N}} \left( | \pmb{n}' \rangle - | \pmb{n}(t) \rangle \right) \cdot \zeta_{\pmb{n}',\pmb{n}}, \\
\! \! \! \! \! \! \! \! \!  &=&  \sum_{ q \in [r]} \left( | \pmb{n} - \pmb{\Delta}_q \rangle - | \pmb{n}(t) \rangle \right) \cdot \zeta_{q,\pmb{n}}, \nonumber \\
\zeta_{q,\pmb{n}} &=& \zeta_{\pmb{n}-\pmb{\Delta}_q,\pmb{n}}
\eea
where $|\pmb{n} \rangle = (|n_1 \rangle, .., |n_s \rangle)^T$. We note that most terms annul, except for those where the difference $\pmb{n}'-\pmb{n} = \pmb{\Delta}_q$ exactly corresponds to the outcome of a reaction $q \in [r] = \{1, 2, ..,r\}$, which for now are assumed to each be first-order reactions in terms of a single species. We denote $h(q)$ the index of the reactant species in reaction $q$ (i.e. 
$\mathbb{S}^{\ominus}_{h(q),q} = 1$), so that
\bea
\langle \zeta_{q,\pmb{n}} \rangle = \kappa_q n_{h(q)} = w_{q,\pmb{n}}  \\
w_{q,\pmb{n}-\pmb{\Delta}_q} = \kappa_q \left(n_{h(q)} + 1 \right)
\eea

We can immediately follow the original derivation of the master equation and obtain
\bea
\! \! \! \! \! \! \! \! \! d_t P_{\pmb{n}} &=& \sum_{\pmb{n}' \in \pmb{N}} \left(P_{\pmb{n}'} w_{\pmb{n}' \rightarrow \pmb{n} } - P_{\pmb{n}}  w_{\pmb{n} \rightarrow \pmb{n}'} \right) \\
\! \! \! \! \! \! \! \! \! &=& \sum_{q \in [r]} 
\left(P_{\pmb{n}-\pmb{\Delta}_q} w_{q, \pmb{n}-\pmb{\Delta}_q} - P_{\pmb{n}}  w_{q, \pmb{n}} \right) \nonumber
\eea
by conservation of probability, the total sum is conserved. 
\bea
\! \! \! \! \! \! \! \! \!  \sum_{\pmb{n}} d_t P_{\pmb{n}} &=& \sum_{q \in [r], \pmb{n}} \left(P_{\pmb{n}-\pmb{\Delta}_q} w_{q, \pmb{n}-\pmb{\Delta}_q} - P_{\pmb{n}}  w_{q, \pmb{n}} \right) \nonumber \\
&=& 0 
\eea
It can readily be shown that equality of the summation holds for each reaction individually\footnote{for instance by setting all other rates than $\kappa_q$ to 0}
\bea
\! \! \! \! \! \! \! \! \! \sum_{ \pmb{n}} P_{\pmb{n}-\pmb{\Delta}_q} w_{q, \pmb{n}-\pmb{\Delta}_q} &=& \sum_{ \pmb{n}} P_{\pmb{n}} w_{q, \pmb{n}} \nonumber \\
\! \! \! \! \! \! \! \! \! \sum_{ \pmb{n}} P_{\pmb{n}-\pmb{\Delta}_q} \kappa_q  \left(n_{h(q)} + 1 \right) &=& \sum_{ \pmb{n}} P_{\pmb{n}} \kappa_q n_{h(q)}  \nonumber
\eea
Note that terms where $n_{h(q)} = 0$ do not count towards the average on the right hand side, and corresponds to states where $X_{h(q)}$ cannot react. These terms are automatically avoided in the summation $\pmb{n} - \Delta_q$, which shifts each term to its corresponding term obtained by one instance of reaction $q$. In infinite summations, it thus directly follows that both sums contain the same nonzero terms. The case of finite summation occurs if $n_{h(q)}$ has some maximum value $n^*$. In this instance, the left hand side sum will contain terms corresponding to $n_{h(q)}=n^*+1$, which correspond to states that cannot be occupied. In either scenario the two sides contain the exact same terms. More generally, we thus have for any vector $\pmb{n}$ 
\be
\! \! \! \! \! \! \! \! \! \sum_{ \pmb{n}} P_{\pmb{n}-\pmb{\Delta}_q} \left(\pmb{n}-\pmb{\Delta}_q\right) = \sum_{ \pmb{n}} P_{\pmb{n}} \pmb{n} \nonumber
\ee
By the termwise correspondence, it follows that higher moments follow the same cancellation
\be
\! \! \! \! \! \! \! \! \! \sum_{ \pmb{n}} P_{\pmb{n}-\pmb{\Delta}_q} \kappa_q  \left(n_{h(q)} + 1 \right)^2 = \sum_{ \pmb{n}} P_{\pmb{n}} \kappa_q n_{h(q)}^2  \nonumber
\ee
We now define an expectation value 
\be
\langle n_k \rangle = \sum_{\pmb{n}} n_k P_{\pmb{n}}
\ee
and now
\bea
\! \! \! \! \! \! \! \! \! d_t \langle \pmb{n} \rangle = \sum_{\pmb{n}} \left(\pmb{n} d_t P_{\pmb{n}}\right) \\
\! \! \! \! \! \! \! \! \! = \sum_{q \in [r], \pmb{n}} \pmb{n} \left(P_{\pmb{n}-\pmb{\Delta}_q} w_{q, \pmb{n}-\pmb{\Delta}_q} - P_{\pmb{n}}  w_{q, \pmb{n}} \right) \nonumber 
\eea
We can rewrite each term in this sum as
\bea
\! \! \! \! \! \! \! \! \! (\pmb{n}-\pmb{\Delta}_q) P_{\pmb{n}-\pmb{\Delta}_q} w_{q, \pmb{n}-\pmb{\Delta}_q} - \pmb{n} P_{\pmb{n}}  w_{q, \pmb{n}} \\
\! \! \! \! \! \! \! \! \! + \pmb{\Delta}_q  P_{\pmb{n}-\pmb{\Delta}_q} w_{q, \pmb{n}-\pmb{\Delta}_q} \nonumber \
\eea
And by termwise correspondence for means and higher moments, we are left with
\bea
\! \! \! \! \! \! \! \! \! d_t \langle \pmb{n} \rangle &=& \sum_{q \in [r],\pmb{n}} \pmb{\Delta}_q  P_{\pmb{n}-\pmb{\Delta}_q} \kappa_q \left(n_{h(q)}+1\right) \\
&=& \sum_{q \in [r]} \pmb{\Delta}_q \kappa_q \langle n_{h(q)} \rangle  \nonumber \  
\eea
Now, defining 
\bea
\! \! \! \! \! \! \! \! \!  \mathbb{S} &=& \left( \pmb{\Delta}_1, .., \pmb{\Delta}_r \right) \\
\! \! \! \! \! \! \! \! \!  \pmb{J}(\langle \pmb{n} \rangle)  &=& \left( \kappa_1 \langle n_{h(1)} \rangle, .., \kappa_r \langle n_{h(r)} \rangle \right) \\
\! \! \! \! \! \! \! \! \!  &=& K  \langle \pmb{n} \rangle  \nonumber  \\ 
\! \! \! \! \! \! \! \! \!  K_{q,h(q)} &=& \kappa_{q}, \ \ \ K_{q,j \neq h(q)} = 0  
\eea
we then obtain
\be
d_t \langle \pmb{n} \rangle = \mathbb{S} \pmb{J}(\langle \pmb{n} \rangle) = \mathbb{S}  K  \langle \pmb{n} \rangle 
\ee
from which we conclude that from a Markov process involving population with first-order transitions in that populations, an ODE representation in terms of population means is obtained that follows the CRN description. 

Merging collinear transitions can from hereon be carried out as in the main text to obtain 
\be
d_t \langle \pmb{n} \rangle = \mathbb{S} \pmb{J}(\langle \pmb{n} \rangle) = \mathbb{S}^* \pmb{J}^*(\langle \pmb{n} \rangle) \label{equation:meaneq}
\ee
Left nullvectors of $\mathbb{S}^*$ now correspond to constraints on mean populations. Note that if the only possible populations are 0 or 1, the mean population is simply the occupation probability, and Eq. \eqref{equation:meaneq} then becomes the master equation Eq. \eqref{equation:mastereq}.

\end{document}